\documentclass[manuscript]{acmart}

\AtBeginDocument{%
  }

\setcopyright{acmlicensed}
\copyrightyear{2026}
\acmYear{2026}
\acmDOI{XXXXXXX.XXXXXXX}
\acmConference[CHI '26]{CHI Conference on Human Factors in Computing Systems}{April 13--17,
  2026}{Barcelona, Spain}

\acmISBN{978-1-4503-XXXX-X/2018/06}

\author{Ameemah Humayun}
\authornote{Both authors contributed equally to this research.}
\affiliation{%
  \institution{Lahore University of Management Sciences}
  \city{Lahore}
  \country{Pakistan}}
\email{26100175@lums.edu.pk}

\author{Bushra Zubair}
\authornotemark[1]
\affiliation{%
  \institution{Lahore University of Management Sciences}
  \city{Lahore}
  \country{Pakistan}}
\email{26100006@lums.edu.pk}

\author{Maryam Mustafa}
\affiliation{%
  \institution{Lahore University of Management Sciences}
  \city{Lahore}
  \country{Pakistan}}
\email{maryam_mustafa@lums.edu.pk}

\begin{document}

\title{Between Myths and Metaphors: Rethinking LLMs for SRH in Conservative Contexts}




\begin{abstract}
Low-resource countries represent over 90\% of maternal deaths, with Pakistan among the top four countries contributing nearly half in 2023. Since these deaths are mostly preventable, large language models (LLMs) can help address this crisis by automating health communication and risk assessment. However, sexual and reproductive health (SRH) communication in conservative contexts often relies on indirect language that obscures meaning, complicating LLM-based interventions. We conduct a two-stage study in Pakistan: (1) analyzing data from clinical observations, interviews, and focus groups with clinicians and patients, and (2) evaluating the interpretive capabilities of five popular LLMs on this data. Our analysis identifies two axes of communication (referential domain and expression approach) and shows LLMs struggle with semantic drift, myths, and polysemy in clinical interactions. We contribute: (1) empirical themes in SRH communication, (2) a categorization framework for indirect communication, (3) evaluation of LLM performance, and (4) design recommendations for culturally-situated SRH communication.
\end{abstract}



\received{20 February 2007}
\received[revised]{12 March 2009}
\received[accepted]{5 June 2009}

\maketitle

\section{INTRODUCTION}
In this study, we investigate how indirect communication strategies in maternal health settings challenge LLM-based interventions in low-income, low-resource countries. Globally, these settings account for more than 9 out of 10 maternal deaths -- most of which are preventable~\cite{WHO2025Maternal}. On average, one in 289 births in low-income countries results in a maternal death, compared with just one in 10,000 in high-income countries~\cite{WHO2025Maternal}. Within this global disparity, Pakistan stands out as one of four countries that together accounted for nearly half of all maternal deaths in 2023;  according to the World Health Organization (WHO), Pakistan recorded 11,000 maternal deaths that year, contributing a staggering 4.1\% to the global total~\cite{APP2025Dawn}. This is despite the modest improvements in recent years, with the country's maternal mortality ratio (MMR) still at an unsettling 155 per 100,000 live births in 2023~\cite{WHO2025Pakistan} -- far above the single-digit MMRs seen in much of the developed Global North~\cite{WorldBank2025MMR}. With 27 preventable maternal deaths per day, Pakistan is far from achieving the Sustainable Development Goals (SDGs) promised in 2015~\cite{WHO2025Pakistan}.

While the HCI community cannot directly address structural and policy barriers, it plays a critical role in designing technologies that improve communication, access, and patient engagement~\cite{Sadeghi2024HCI, Yeung2023Digital, Maeda2025Digital}. This is particularly important in low-resource contexts, where the majority of maternal deaths are due to preventable complications and can be addressed through the timely identification of risk factors~\cite{WHO2025Maternal}. Since these contexts have the additional burden of low consultation time per patient~\cite{Irvinge017902}, the rushed environments often lead to incomplete histories and missed warning signs. Hence, the automation of the initial screening to flag high-risk cases may significantly reduce the global number of maternal deaths. In the last year, large language models (LLMs) have shown great promise for supporting and improving healthcare delivery~\cite{Nazi2024Large, Lawrence2024Opportunities}, demonstrating strong performance in tasks vital to such effective health interventions, including extracting and synthesizing medical knowledge, answering complex health-related queries, generating patient-friendly explanations, and supporting context-aware decision-making~\cite{Aydin2024PatientEducation, Umerenkov2023Deciphering, Garcia2024Extraction, Singhal2023Knowledge}. 

However, there are certain limitations to the efficacy of these interventions. AI suffers through several biases, including those of a racial, gender, and ethnic nature~\cite{Navigli2023BiasesLLMs, Sap2019RiskRacialBias, OConnor2024GenderBiasAI, Ahn2021MitigatingBERTBias}. More critically, for Global South contexts, LLMs often claim to be language-agnostic but seem to be predominantly English-centric and hence inherit Western biases in language as well~\cite{Bender2011LanguageIndependenceNLP, Schut2025MultilingualLLMs}. Hence, the effectiveness of applying these tools in underserved contexts across the Global South is particularly constrained due to the disparity in both the availability of language datasets~\cite{Silva2024BenchmarkingLowResourceMT, Bender2011LanguageIndependenceNLP, Joshi2020LinguisticDiversityNLP} and in LLMs' ability to navigate cultural nuances in these settings~\cite{Deva2025Kya}.

These barriers become especially problematic when designing health-tech interventions for localized contexts; effective clinician-patient communication is central to quality healthcare~\cite{KELDER2022858}, making the faithful replication of this communication one of the core challenges in this domain. LLMs often struggle with nuanced conversations in non-English languages, particularly in specialized and socially complex contexts like sexual and reproductive health (SRH)~\cite{Deva2025Kya}. Because SRH is often taboo and inappropriate to discuss openly, the stigma and shame surrounding these topics necessitate euphemistic language, even in healthcare settings. Additionally, local vernacular may often be used to describe SRH concerns instead of English terms or even formal terminology within the local language. Consequently, conversational agents may successfully handle surface-level exchanges but often fail in such culturally and linguistically nuanced situations.

In this paper, we examine the intersection of LLMs and the use of indirect communication in SRH. While prior work has addressed cultural sensitivity in chatbot design and highlighted the scarcity of language resources for AI in low-resource settings, little research has examined how the use of euphemisms and other forms of indirect language impacts the effectiveness of LLMs, especially when deployed for health. To explore this challenge, we evaluate LLM performance on a data sample of indirect SRH language used in low-resource contexts through a two-stage study. First, we present a qualitative investigation including interviews, focus groups, and observation sessions to identify and categorize euphemistic and indirect expressions used by women in Pakistan to discuss SRH concerns. We also identify key challenges in clinician-patient communication that warrant consideration when designing interventions. Second, we perform a performance evaluation of multiple LLMs to interpret these expressions, including LLaMA 3.2, Gemma 3, GPT-OSS, GPT-4o, and Claude Sonnet 4. This mixed approach allows us to connect the lived linguistic realities of healthcare communication with the technical performance limits of LLMs. By understanding women's communication strategies in healthcare settings, we help prioritize design that adapts to existing patterns rather than requiring behavioral change from vulnerable users. Our data sample focuses on Roman Urdu (Urdu written in the Latin script), since it is the most widely used written form of Urdu in Pakistan’s digital communication practices, including SMS, WhatsApp, and social media~\cite{bilal2018analysing}. 

We answer the following three research questions:

\textbf{RQ1:} What types of indirect communication strategies do women in low-income, low-resource contexts like Pakistan use to talk about their maternal health and related SRH concerns, and how can these be systematically categorized?

\textbf{RQ2:} How well do large language models interpret indirect expressions related to SRH in low-income, low-resource contexts, as examined through a case study in Pakistan?

\textbf{RQ3:} Based on the communication strategies identified and the evaluation of LLM performance, what design recommendations can be made to improve LLM-based SRH interventions in low-resource contexts?

In this paper, we develop a systematic framework for categorizing indirect SRH health communication along two axes and evaluate LLM performance against this linguistic reality. We present our two-stage methodology and findings, which reveal that women's communication follows distinct contextual and expressive patterns that challenge current LLMs differently. We also identify key challenges in clinician-patient communication that warrant consideration when designing interventions. Our work extends beyond performance evaluation to propose design principles for leveraging LLMs for healthcare that can operate effectively within existing cultural communication practices rather than requiring users to change their language patterns.

By linking grounded linguistic insights to LLM evaluation, this work advances the design of health technologies that can operate more effectively in culturally complex, resource-constrained environments. In doing so, it contributes to ongoing conversations in HCI about equity, inclusivity, and the role of sociotechnical systems in advancing global health goals.

\section{RELATED WORK}

\subsection{AI-Driven Healthcare Interventions}

Early mobile health (mHealth) interventions in low-resource contexts relied heavily on SMS-based systems for reminders, maternal care follow-ups, and health education. While these interventions proved that digital communication could extend healthcare reach, they were often constrained by low engagement and limited personalization. Labrique et al.~\cite{Labrique} established one of the first systematic frameworks for mHealth, identifying twelve common applications (e.g., client education, provider training, data capture) and situating them within broader health system strengthening. Early projects, such as Baby+ in Pakistan~\cite{BabyPlus}, demonstrated the potential of localized pregnancy apps, but also revealed barriers of usability and sustained adoption in low-literacy populations. Similarly, WhatsApp-based supervision of community health workers in Kenya improved communication flows~\cite{kenyawhatsapp}, yet depended on existing infrastructure and informal practices rather than scalable design.  

As mobile infrastructures matured, conversational agents emerged as more interactive alternatives to SMS. Chatbots like MANDY in primary care~\cite{MANDY} and FeedPal for breastfeeding education~\cite{FeedPal} showed that natural language interaction could improve accessibility and cultural relevance. 

In sexual and reproductive health (SRH), chatbots have expanded access to sensitive information in underserved communities. Initiatives such as Myna Bolo in Indian slums~\cite{MynaBolo}, Ipas’s suite of culturally adapted SRH bots~\cite{Ipas}, SnehAI for Indian adolescents~\cite{SnehAI}, and askNivi across Kenya, India, and Nigeria~\cite{askNivi} illustrated that chatbots can deliver stigma-reducing, localized guidance at scale. Yet, these projects also highlighted persistent barriers, including digital literacy gaps, infrastructural limitations, and enduring stigma around SRH communication. Collectively, these findings suggest that while chatbots are valuable for information dissemination, they struggle to achieve sustained behavioral change without deeper integration into health systems~\cite{RealistSynthesis}.  

The advent of large language models (LLMs) has opened new directions but introduced new risks. Systems like ASHABot for community health workers~\cite{ASHABot}, Socheton in Bangladesh~\cite{Socheton}, Myna Bolo in India~\cite{Deva2025Kya}, and Awaz-e-Sehat in Pakistan~\cite{Awaz-e-Sehat} demonstrated that LLMs could handle more complex, context-sensitive queries, extending beyond the capabilities of rule-based systems. However, HCI researchers caution that these interventions remain under-evaluated in real-world contexts. Key concerns include bias in training data, overreliance on AI over human expertise, and the neglect of frontline health workers’ perspectives in scaling interventions~\cite{Jo2023,AlGhadban2023,Karusala2023}. Reviews of African deployments further stress that enthusiasm for AI far outpaces evidence of its actual health impact~\cite{Phiri2023}.  
Despite this progress, the evidence consistently highlights limitations in scalability, contextual and cultural adaptability, and rigorous evaluation. While proof-of-concept deployments are important, it is necessary to systematically assess how chatbots--particularly those powered by LLMs—-can be equitably integrated into health systems in accordance with the norms, culture, and colloquialisms present, and to address infrastructural and literacy barriers to build adequate trust in sensitive domains such as maternal and reproductive health.

\subsection{Indirect Communication in Healthcare}

The foundations of indirect communication in healthcare settings lie in (1) the navigation of social taboos and (2) a lack of health literacy (HL); either individuals feel they are not allowed to say the required words, or they simply do not know them.

\subsubsection{Navigating Taboos} Chaudhri et al. analyzed thirteen English articles to study the interplay of euphemisms and taboos in Pakistani society. They noted that in some households, the shame associated with something as natural as menstruation is so strong that not only are women expected to avoid openly speaking about their cycles, but may even go so far as to fake fasts during the Islamic month of Ramadan -- despite the religious exemption for menstruating women -- lest the men in their family notice and conclude that they are menstruating~\cite{Chaudhri2022Taboos}. A study in Kenya studied several interactions between a female doctor and male patients discussing their reproductive health to showcase how even men faced the discomfort of discussing taboo subjects and employed indirect strategies like euphemisms, silences, and gestures. They highlighted how these clinician-patient conversations were highly dependent on the healthcare provider's ability to read between the lines and work with indirect strategies rather than against, and that often even the doctor felt the weight of social taboos during consultations. ~\cite{Ouma2025NavigatingTaboosMensHealth}. A systematic review by Kelder et al. identified seven key communication strategies in sexual health consultations: avoiding delicate terms, delaying potentially sensitive words, using assumptive talk, providing generalized advice, deploying patients' own language choices, depersonalization, and allowing patient-initiated advice~\cite{KELDER2022858}. These highlight the complex nature of the way sensitive and taboo issues are navigated. The review particularly highlights~\cite{Weijts1993, Silverman1991} when it comes to avoiding and delaying sensitive topics, noting that both parties in a clinician-patient setting systematically avoid explicit sexual terminology through multiple mechanisms, including using words like "it" and "that" to refer to previously introduced delicate topics, which typically does not create interpretation problems since context has been established previously. Healthcare professionals often adopt patients' own omission patterns in a collaborative approach to managing sensitive content~\cite{Weijts1993, Silverman1991}.

\subsubsection{Health Literacy} WHO defines health literacy as "being able to access, understand, appraise and use information and services in ways that promote and maintain good health and well-being...it includes the ability to think critically about, as well as the ability to interact and express personal and societal needs for promoting health."~\cite{who2024healthliteracy}. Several studies in the Global North have noted that the lack thereof causes significant barriers in communication due to restricted medical vocabulary, limited background knowledge, and challenges in processing new information; as a result, they often report difficulties in reporting symptoms and understanding their diagnoses and treatment plans~\cite{Shahid2022LowHealthLiteracyOutcomes, Parker2001HL, hasannejadasl2022healthliteracyeoncologycare}. Healthcare providers note that low HL patients are often unable to articulate their symptoms, understand or adhere to advice and explanations, and even expect healthcare providers to make important decisions for them~\cite{Murugesu2022LowHealthLiteracyCommunication}. However, these extensive studies mostly exist in the Global North. Few have shown how these gaps manifest in low-resource, conservative, non-English contexts, and how exactly they may relate to developing LLM-based interventions, which is the gap we aim to fill.

\subsection{Bridging Sociolinguistic Reality and AI Designs}

Cultural sensitivity can be understood through Deva et al.’s \cite{Deva2025Kya} layered framework spanning linguistic, sociocultural, and systemic levels. Linguistically, local dialects, euphemisms, and code-mixed expressions shape how people articulate health concerns. Socioculturally, stigma, taboos, and family-based decision-making constrain SRH communication. Systemically, power hierarchies and biomedical assumptions structure clinical encounters. This layered view is crucial in SRH, where indirect speech and culturally coded terms are the norm. Even medical advice cannot be separated from its cultural context; Arueyingho et al. \cite{nigeria2024} show that Nigerian patients interpret diabetes through spiritual and collective frames, making Western-oriented advice such as “low-carb cereal” irrelevant to diets centered on yam or cassava. Similar misalignments in SRH—around marriage, modesty, or stigma—undermine trust in AI systems. When cultural nuance is absent, AI reproduces bias. Lee et al. \cite{migrants2024} find that conversational AI misrepresented migrant realities, reinforcing stereotypes. Joshi et al. \cite{joshi2020state} highlight how Western-centric training data sidelines underrepresented dialects, leading to errors such as confusing \textit{rokna} (contraception) with \textit{rukna} (miscarriage) \cite{Deva2025Kya}. These mismatches reveal structural gaps between global AI infrastructures and local practices. Euphemisms, central to SRH discourse, further complicate interpretation. They soften realities, preserve politeness, and shape illness perception. Ezeugo and Chukwu \cite{ezeugo2023} show euphemisms structuring doctor–patient talk, while Tayler and Ogden \cite{tayler2005} find softer terms reduce anxiety but alter beliefs about severity. In SRH, phrases like “missed my date” (menstruation) or “bring a visitor” (pregnancy) are widespread but hard for AI to parse. Zhu et al. \cite{zhu2021euphemism} show euphemism detection is harder than profanity detection, especially with multi-word forms \cite{zhu2021phrase}. Thus, SRH communication depends on exactly the forms AI struggles to interpret. Recent work explores cultural adaptation. Baik et al. \cite{chatbot2024} show that tailoring chatbot style (authoritative, informal, sentimental) improves perceived alignment. Aleem et al. \cite{mentalhealth2024} and Sehgal et al. \cite{india2023} demonstrate that culturally adjusted conversational strategies foster trust in mental health contexts. At a population level, Davies et al. \cite{publichealth2024} synthesize evidence that culturally adapted public health messaging increases engagement and uptake. Yet, few studies address SRH directly. Existing efforts adapt tone or style but rarely engage with the semantic complexity of dialect mixing, taboo terms, or euphemisms--precisely where LLMs potentially falter.  

\section{METHODOLOGY}

\begin{figure*}[t] 
    \centering
    \includegraphics[width=\linewidth]{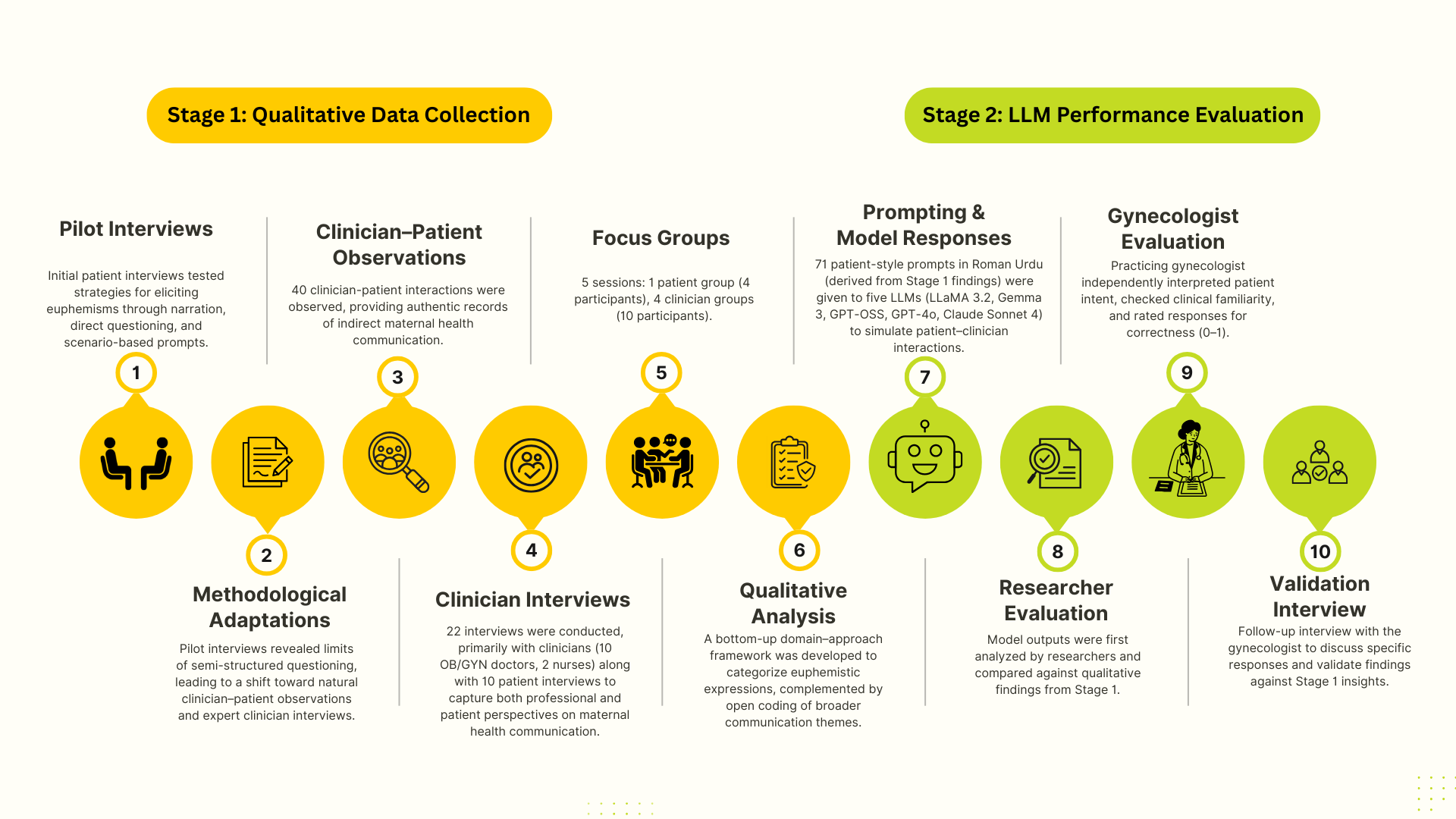}
    \caption{Design overview of our two-stage study conducted in a charitable hospital in Lahore, Pakistan.  
    Stage 1 (Steps \textbf{1–6}, Qualitative Data Collection): patient interviews, clinician–patient observations, focus groups, and clinician interviews informed methodological adaptations and qualitative analysis of euphemistic and indirect SRH communication.  
    Stage 2 (Steps \textbf{7–10}, LLM Performance Evaluation): 71 prompts derived from Stage 1 were tested on five LLMs (LLaMA 3.2, Gemma 3, GPT-OSS, Claude Sonnet 4, GPT-4o). Model responses were analyzed by researchers, rated by a gynecologist for correctness, and validated through follow-up interviews.}
    \label{fig:methodologytimeline}
\end{figure*}

As shown in Figure~\ref{fig:methodologytimeline}, the study was conducted in two stages. The first was a qualitative study to learn common communication patterns in healthcare settings, involving clinicians and patients in Lahore, Pakistan. This included semi-structured interviews, focus groups, and observation sessions. Details can be found in Table~\ref{tab:data_collection}. All interviews were conducted by two female researchers in Urdu. All participants were women and informed consent was obtained. The second stage aimed to test the performance of several LLMs, including open source (Llama3.2, Gemma3, GPT-OSS ) and proprietary (Claude Sonnet 4, GPT 4), using prompts with and without context.

\subsection{Stage 1: Qualitative Data Collection}

\subsubsection{Settings and Access}
Data collection occurred in a charitable hospital. This provided us access to patients with low-to-middle-income backgrounds and formally trained clinicians experienced in communicating with them. We maintained contact with all participating doctors and periodically sent reminders requesting any additional insights they recalled or newly encountered cases. Consistent contact was also maintained for rechecking and clarifying the collected data.

Participants were chosen by convenience sampling. The following tables summarize our data collection methods. 

\begin{table}[h]
\caption{Overview of qualitative data collection methods and participant demographics across four data collection approaches to catalogue indirect communication in SRH settings.}
\label{tab:data_collection}
\begin{tabular}{lc}
\hline
\textbf{Method/Participant Type} & \textbf{Count} \\
\hline
\multicolumn{2}{l}{\textit{\textbf{Primary Data Collection}}} \\
Clinician-Patient Observations & 40 \\
Semi-structured Interviews & 22 \\
Focus Group Sessions & 5 \\
\hline
\multicolumn{2}{l}{\textit{\textbf{Interview Breakdown}}} \\
Patients & 10 \\
OB/GYN Doctors & 10 \\
Nurses & 2 \\
\textbf{Total} & \textbf{22} \\
\hline
\multicolumn{2}{l}{\textit{\textbf{Focus Group Breakdown}}} \\
Patient Groups (4 participants) & 1 \\
Doctor Groups (10 participants) & 4 \\
\textbf{Total (14 participants)} & \textbf{5} \\
\hline
\end{tabular}
\end{table}

\subsubsection{Pilot Study and Methodological Adaptations}
An initial interview protocol for patients was developed collaboratively by the two lead researchers and subsequently reviewed by the Head of the Department of Gynecology at the collaborating hospital. This protocol incorporated three distinct strategies for eliciting indirect language regarding SRH:
\begin{enumerate}
    \item \textbf{Implicit Search for Euphemisms (Narration of Experiences):} Open-ended prompts invited participants to narrate their maternal health experiences without indicating our focus on euphemisms, capturing authentic linguistic patterns through natural storytelling.
    \item \textbf{Explicit and Directed Search for Euphemisms:} Direct questioning asked participants how they communicated specific situations, prompting recall of exact words and expressions used in maternal health discussions.
    \item \textbf{Scenario-Based Questions:} Hypothetical scenarios allowed participants to respond from a detached perspective, potentially reducing hesitation around sensitive topics.
\end{enumerate}

However, pilot patient interviews revealed significant limitations with semi-structured interviews for eliciting euphemistic language and shifted toward observing natural clinician-patient interactions and conducting extended conversations with health professionals. We detail these methodological challenges and their implications in Section~\ref{subsec:pilotissues}. For richer insights, we reduced patient interviews and shifted to observing natural clinician-patient interactions~(with consent) and conducting expert interviews with clinicians who could access broader experiential datasets.

\subsection{Qualitative Analysis} 

We used two distinct analytical processes. First, during interviews and observations, we noted recurring indirect expressions and communication patterns. As new terminologies emerged across participants, we identified commonalities in what was being referenced (domains) and how it was expressed (approaches). This bottom-up pattern recognition led to our systematic domain-approach framework, which we present in greater detail in Section~\ref{sec:framework}. This was then applied consistently across all data to categorize indirect language use in SRH settings. Second, we conducted open coding to identify recurring themes in clinician-patient interactions that should be accounted for when designing health interventions.

\subsection{Stage 2: LLM Performance Evaluation}

We systematically evaluated the interpretive capacity of five large language models using prompts derived from our qualitative data collection. The evaluation included three open-source models accessed via Ollama (Gemma3, LLaMA3.2, and GPT-OSS) and two proprietary models (GPT-4o and Claude Sonnet 4). 

To simulate authentic doctor–patient interactions, all prompts were framed from the patient’s perspective in Roman Urdu, often drawn directly from our qualitative findings. Each model was prompted with the persona of a gynecologist working in a low-resource public hospital in Pakistan. The persona emphasized cultural sensitivity, communication accessible to low-literacy patients, and avoidance of English technical jargon. Models were asked to return both (1) the corresponding medical term and (2) a culturally sensitive, clinician-like response in Roman Urdu. See Table~\ref{tab:llm_input} in the appendix for the exact persona and prompt.

\subsubsection{Data Sample and Prompts}
A total of 71 prompts were used from the original 80 terms collected from our initial qualitative research. Some formally recognized terms (eg, \textit{[egg-container]}, \textit{[child-container]}) were not included in separate prompts but used within other prompts.

\subsubsection{Response Analysis}

Model outputs were first analyzed by researchers in light of our qualitative research findings, then independently evaluated by a practicing gynecologist. For each patient query, the gynecologist (1) interpreted what she believed the patient intended, (2) noted whether she had encountered the term in clinical practice, and (3) identified possible synonyms or alternative expressions. She then rated model responses from the three highest-performing LLMs on correctness/accuracy (0-1 binary scale), while also providing detailed comments for each response. Finally, we conducted a follow-up interview with the gynecologist to discuss specific responses and validate findings against our qualitative knowledge.

\subsection{Notation Conventions}
For readability, Urdu phrases and words are translated literally, italicized, and enclosed within square brackets, followed by a round bracket explanation if necessary: \textit{[translated term]} (medical term in English). In some places, \textit{"local term"} (translation) is also used. Literal translations may sound awkward in English, but are necessary to convey the essence of the collected data and preserve the linguistic patterns observed in participants' communication.

\subsection{Positionality}
All authors are women based in Lahore, Pakistan, with fluency in English, Urdu, and Punjabi. Our lived experiences with local social taboos and euphemistic language enabled nuanced interpretation of participants' communication patterns. While our gender makes us stakeholders in maternal health conversations, we acknowledge that our privileged socioeconomic and educational backgrounds may not fully reflect the experiences of lower-income participants, particularly regarding health literacy.

\subsection{Ethical Considerations}
All participants provided informed consent after the study was explained in Urdu. All interviews were conducted by two lead female researchers to ensure same-gender interactions and avoid violating social taboos around discussing intimate health matters with the opposite gender. Participants were allowed to decline discussing any topics they found uncomfortable and were informed they could withdraw from the study at any time. All interactions were within hospital settings where participants already felt comfortable discussing health matters. We obtained consent to record interviews, which were then transcribed and anonymized with identifying information removed before analysis. When recording was not possible, detailed notes were taken. For observations, we obtained consent from both parties and positioned ourselves to document communication patterns without interfering with clinical care.

\section{FINDINGS}

We present our findings in three sections: (1) First, we detail our framework, presenting the five indirect communication approaches and five referential domains that emerged from our analysis. (2) Then, we examine broader patterns in clinician-patient interactions, including communication breakdowns and information gatekeeping behaviors. (3) Finally, we evaluate how well large language models can interpret these indirect communication strategies. The first two of these sections map onto RQ1, the third maps onto RQ2, and our discussion in Section~\ref{sec:disc} maps onto RQ3.

\section*{CATEGORIZATION FRAMEWORK FOR INDIRECT COMMUNICATION IN SRH}\label{sec:framework}

\begin{figure*}[t]
    \centering
    \includegraphics[width=\linewidth]{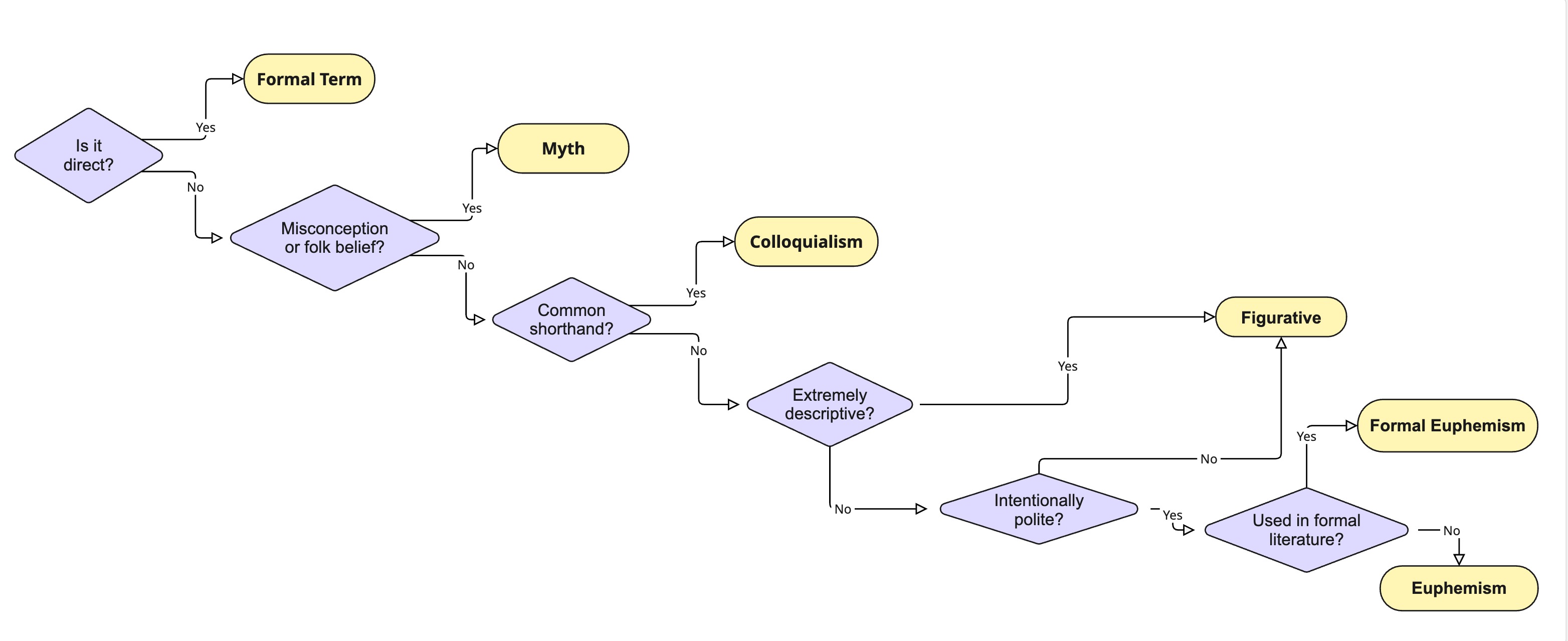}
    \caption{Categorization framework for communication in sexual and reproductive health (SRH). 
    Decision tree for categorizing sexual and reproductive health (SRH) communication strategies along the \textbf{Communicative Approach} axis. The framework moves from direct, formal terms toward a range of indirect approaches—colloquialisms, euphemisms, figurative expressions, and myths—depending on how a concept is expressed. Each category reflects a different communicative function: convenience (colloquial shorthand), politeness and taboo navigation (euphemism), descriptive compensation (figurative terms), or culturally embedded misconceptions (myths). The diagram shows how everyday language departs from biomedical vocabulary, creating systematic interpretive challenges for both clinicians and LLM-based systems.}
    \label{fig:mapp}
\end{figure*}

In this section, we present a \textbf{Domain-Approach} framework to categorize communication strategies along two axes: (1) referential \textbf{domain} and (2) communicative \textbf{approach}. We define indirect communication strategies as communicative approaches that deviate from \textbf{formal} clinical terminology, i.e., direct translations of medical terms found in clinical literature, textbooks, and official medical records. While formal approaches use precise biomedical vocabulary, indirect approaches employ alternative methods to navigate social, cultural, or educational barriers in healthcare communication.

\paragraph{\textbf{(1) Referential Domains}} (represent what is being discussed):
\begin{itemize}
  \item \textit{Symptom}: experienced states (e.g., pain, discharge, leaking) (Table~\ref{tab:symptoms_terms})
  \item \textit{Procedure}: surgeries, tests, or interventions (e.g., episiotomy, C-section, UPT) (Table~\ref{tab:procedures_terms})
  \item \textit{Condition}: named diagnoses or findings (e.g., fibroid, ectopic pregnancy) (Table~\ref{tab:conditions_terms})
  \item \textit{Process}: actions or physiological states (e.g., sexual intercourse, pregnancy, breastfeeding) (Table~\ref{tab:processes_terms})
  \item \textit{Anatomical}: body parts (e.g., vagina, uterus, breasts) (Table~\ref{tab:anatomical_terms})
\end{itemize}

\paragraph{\textbf{(2) Communicative Approaches}} (represent how medical concepts are expressed through indirect means):

\begin{itemize}
  \item \textit{Euphemisms}: Politeness strategies for sensitive topics
  \item \textit{Colloquialisms}: Standardized informal medical shorthand
  \item \textit{Figurative}: Descriptive metaphors compensating for limited vocabulary
  \item \textit{Myths}: Folk beliefs and misconceptions shaping health understanding
  \item \textit{Gestures}: Non-verbal and deictic communication
\end{itemize}

These approaches often involve multilingual code-switching (Urdu-English-Punjabi mixing) and can overlap -- for example, terms may be both figurative and euphemistic. When categorizing overlapping cases, we prioritize the dominant communicative function: euphemism for taboo circumvention, colloquialism for neutral shorthand, and figurative for vocabulary compensation. Myth classification overrides other categories when folk beliefs are involved. Figure~\ref{fig:mapp} shows our classification decision tree for determining communicative approaches. The following subsections examine each communicative approach in detail.

\begin{table*}[t]
\centering
\caption{Approach–Domain summary of indirect SRH expressions (concise examples consistent with Appendix Tables~\ref{tab:termsfig}–\ref{tab:myths_terms}). The table synthesizes representative examples across figurative, colloquial, euphemistic, and mythic forms, illustrating how communicative strategies cut across biomedical domains and complicate one-to-one mappings in health technology design. For each approach, we also provide what "sub-type" the examples fall into, which are further elaborated in the respective sections. For general examples, the name of the approach is listed.}

\label{tab:approach_domain_summary}
\small
\setlength{\tabcolsep}{4pt}
\renewcommand{\arraystretch}{1.2}
\resizebox{\textwidth}{!}{%
\begin{tabular}{p{2.8cm}p{2.8cm}p{3.2cm}p{3.2cm}p{4.2cm}}
\toprule
\textbf{Sub-type} & \textbf{Referential Domain} & \textbf{Clinical Concept} & \textbf{Local Term} & \textbf{Literal Translation / Notes} \\
\midrule
\multicolumn{5}{c}{\textbf{FIGURATIVE}} \\
\midrule
Hyper-descriptive & Symptom & Leukorrhea & \textit{charbi} & (stringy pieces of) fat \\
Vague & Symptom   & Vaginal leaking/Urinary Incontinence  & \textit{shalwar bheeg gayi}   & shalwar soaked; indicates volume \\
Valid polysmey & Symptom   & Menstrual clot, miscarriage, prolapse   & \textit{gosht ka piece}                 & piece of meat \\
\midrule
\multicolumn{5}{c}{\textbf{COLLOQUIAL}} \\
\midrule
Colloquialism & Condition & Gestational diabetes & \textit{baby sugar}    & everyday shorthand \\
Semantic Drift & Process & Vaginal Delivery & \textit{normal}         & everyday shorthand \\
Semantic Drift & Procedure & Episiotomy & \textit{chhota operation}     & small operation (minor surgery) \\
\midrule
\multicolumn{5}{c}{\textbf{EUPHEMISM}} \\
\midrule
Euphemism & Anatomical & Vagina                  & \textit{neechey wali jaga}  & area down there \\
Euphemism & Process    & Sexual intercourse      & \textit{shohar sey miltay waqt} & when meeting the husband \\
Euphemism & Process    & Pregnancy / late menses & \textit{tareekh nahi ayi}   & date has not come \\
\midrule
\multicolumn{5}{c}{\textbf{MYTH}} \\
\midrule
Misconception   & Symptom   & Vaginal discharge            & \textit{haddi}   & bones dissolving (misconception) \\
Folk Belief   & Condition & Recurrent miscarriages       & \textit{athra}   & curse of miscarriages (folk belief) \\
Folk Polysemy & Condition & Fibroid / cyst umbrella term & \textit{rasoli}  & umbrella term for growths \\
\bottomrule
\end{tabular}}
\end{table*}

\subsection{Colloquialisms}
Colloquialisms (Table~\ref{tab:termscolo}) represent neutral shorthand used in routine practice (e.g., \textit{[big] operation} for C-section). We use "colloquialisms" to refer to everyday vernacular terms that do not necessarily stem from cultural taboos but represent a standardized informal medical language within our specific healthcare context, used predominantly during clinician-patient interactions. Unlike euphemisms, the motivation for this communicative approach is convenience rather than politeness. We classified terms as colloquial when: (1) participants used them matter-of-factly without hesitation or circumlocution, (2) both patients and providers demonstrated mutual understanding without explanation, and (3) doctors self-reported them as common ways of referral when speaking to patients in day-to-day interactions.

These terms appeared predominantly when discussing procedures and conditions, where "procedures" encompass surgeries, tests, and medical interventions. This communicative approach exhibited the most code-switching, frequently borrowing English terminology. We observed that common procedures had dedicated colloquial phrases or directly adopted English terms (e.g., "PID"), while less familiar conditions were explained in detail. This category also displayed the most semantic drift:

\paragraph{\textbf{Semantic Drift}}\label{subsub:semdrift}
Many colloquial terms appear vague, but actually have specific definitions within OB/GYN settings. For instance, \textit{"chota operation"} (small operation) is commonly used to refer to minor surgeries in Pakistan, but specifically denotes episiotomy within obstetric contexts. Similarly, \textit{"bara operation"} (big operation), commonly used for major surgeries, refers specifically to C-sections in obstetric contexts. These colloquialisms exemplify semantic drift within specialized medical communities i.e. evolving from broad categorical terms (major/minor surgery) to highly specific procedural references (C-section/episiotomy) within obstetric contexts. This drift occurs through repeated usage patterns where the general term becomes specifically associated with the most common procedure in that category; as obstetric clinicians and patients repeatedly encounter episiotomies as the most frequent "small operation" and C-sections as the most common "major operation,", it creates a specialized lexicon that facilitates rapid communication among insiders while potentially excluding outsiders unfamiliar with the semantic shift -- and sometimes causing insiders confusion too, as explained in Section~\ref{subsub:miscom}.

\begin{quote}\small\itshape
\textbf{Interviewer:} All interviews we've had with doctors so far mention [small operation] is commonly used for episiotomy. Can you clarify whether this is always specifically used for episiotomy in OB/GYN contexts, or is it just sometimes the case? Because of course we use the term [small operation] commonly for any minor surgery.\\
\textbf{Participant:} Yes, in OB/GYN we only call episiotomy [small operation]. Nothing else.\\[2pt]
\textemdash~\textbf{P8, Doctor}
\end{quote}

\subsection{Euphemisms}

Euphemisms (Table~\ref{tab:euphemistic_terms}) represent politeness strategies to navigate social taboos (e.g., \textit{[meeting husband]} for sexual intercourse). Given the sensitive nature of SRH topics, most collected terminology has euphemistic undertones, but we include here only those primarily motivated by social taboo avoidance rather than vocabulary limitations or clinical efficiency.

These terms focus predominantly on anatomy and processes closely related to sexuality and reproduction. Euphemisms are often the most recurring expressions in SRH conversations and form the foundation for compound medical inquiries—for example, post-coital symptoms are routinely assessed through questions like \textit{[does it hurt or bleed after meeting your husband?]}. We observed multiple variations for sexual intercourse, with most expressions centering on "meeting" the husband.

This pattern of linguistic avoidance extends beyond sexual terminology to pregnancy itself. When discussing pregnancy with family members, participants demonstrated how euphemistic language becomes so embedded that direct terminology feels inappropriate. Religious framing emerged as a common strategy, where pregnancy is discussed through spiritual rather than biological terms. One participant consistently avoided direct pregnancy terminology, instead framing conception as divine blessing:

\begin{quote}\small\itshape
\textbf{Participant:} In this there are no words, just happiness that Allah has blessed us with this joy. There are no words that I can describe.\\
\textbf{Interviewer:} When you told your mother or mother-in-law, what words did you use to talk to them?\\
\textbf{Participant:} Everyone was very happy and just said that those who don't have [children], may He give to them too. These are our words and until our end these are our words - that those who don't have, may Allah give to them too, and what we asked for, He gave.\\
\textbf{Interviewer:} When you told [someone] again, what words did you use?\\
\textbf{Participant:} Again I don't have words, only happiness is felt. That cannot be described in words.\\[2pt]
\textemdash~\textbf{P3, Patient}
\end{quote}

Euphemisms can be particularly confusing when conflated with other concepts. For example, \textit{"pishaab wali jaga"} (urine area) is sometimes used euphemistically to refer to the vagina, and at other times literally for the urinary tract.

\subsection{Figurative}

\begin{quote}\small\itshape
There are some words that we say in different ways. We and [the undereducated patients] are different. Because we are educated, we can make each other understand. But they can't.\\[2pt]
\textemdash~\textbf{P12, Nurse}
\end{quote}

The figurative approach (Table~\ref{tab:termsfig}) includes terminology most closely associated with low health literacy, where patients create ad-hoc descriptors to compensate for limited biomedical vocabulary (e.g., \textit{[inside is falling out]} for vaginal prolapse). This approach exists in extremes: either (1) vague indications (e.g., \textit{[pain under abdomen]} for pelvic pressure) or (2) hyper-descriptive metaphors (e.g., \textit{[stringy pieces of fat]} for white vaginal discharge). These expressions reflect patients grasping for words to accurately convey their symptoms, often ending up with either too little or too much descriptive detail. While some overlap with euphemisms due to politeness requirements, the figurative quality is clearly dominant.

Hyper-descriptive figurative terms tend to be creative, rare, and highly individualistic. While some become standardized over time (e.g., \textit{[chunk of meat]} for menstrual clot), others remain exceptionally rare. The term \textit{[fat]} for leukorrhea exemplifies this rarity—reported by two doctors (P8. P20) but unrecognized by others, yet showing remarkable descriptive accuracy. While LLMs and online images typically represent \textit{"charbi"} as chunky pieces of fat, the participant meant sinewy, string-like strands of fat similar to what comes off when cleaning meat, reflecting visual similarity to actual vaginal discharge. This demonstrates how figurative language can be highly individualized and requires both contextual and cultural knowledge to interpret accurately.

Vague terms, on the other hand, create an entirely new challenge. Here, patients may describe a concept so vaguely that their descriptor becomes polysemic. We identify two distinct categories of polysemy in healthcare communication: (1) \textbf{valid polysemy}, where \textit{general descriptors} appropriately cover multiple similar medical concepts, and (2) \textbf{folk polysemy}, where patients incorrectly assume \textit{medically specific terms} can generally encompass medically distinct phenomena. Folk polysmey more appropriately falls under Myths and Misconceptions, which we describe in the next section. (Section~\ref{subsub:folkpoly}). The consequences of these are unpacked in Section~\ref{subsub:miscom}.

\paragraph{\textbf{Valid Polysemy}}\label{subsub:validpoly}
Several vague figurative terms demonstrate legitimate polysemic usage where a single, general descriptor appropriately covers multiple related medical concepts. For instance, \textit{"taanka"} (stitch) is used for copper IUD insertion, surgical sutures, and bilateral tubal ligation -- all procedures involving "stitching" anatomical structures. Similarly, \textit{"challa rakhwaya"} (placed a ring inside) applies to both pessaries and IUDs, since both are inserted medical devices. The term \textit{"paani ki thaeli"} (water sac) encompasses both hydrosalpinx and ovarian cysts, both being fluid-filled structures. Pain-related terminology shows similar patterns: \textit{"peedan"} (pains) covers both extreme menstrual pain and labor pains, while \textit{"gosht ka piece"} (piece of meat) describes endometrial tissue passage, menstrual clots, miscarriage tissue, and polyps -- all involving tissue expulsion.

\subsection{Myths and Misconceptions}
We classified expressions as myths when they reflected non-medical belief systems or medically inaccurate explanations that shaped explanation and help-seeking behaviors, regardless of whether they were used directly or had qualities of any other approach. These are either (1) folk beliefs or (2) general misconceptions, and (3) \textit{folk polysemy}, where everyday terms acquire additional, culturally specific meanings that diverge from biomedical interpretation. We identify primary examples for each category:

\paragraph{\textbf{Folk Beliefs}} \textit{Athra} refers to a concept described as a "curse of miscarriages," and is often viewed as contagious. The belief suggests that women can develop this condition after experiencing repeated miscarriages or even after being around someone who has had a miscarriage. During a focus group discussion, a doctor reported: \textit{"A patient came to me and said I think I have athra because my neighbour has it"}. In one focus group, this example sparked discussion among the medical professionals, as the newer doctor had never heard of the term while the experienced doctor confirmed it was quite common among undereducated patients. One doctor reported that a woman wanted to get an abortion (a sensitive subject in Pakistan) because a \textit{Pir} (spiritual leader) warned her that it was necessary as this particular pregnancy would induce \textit{Athra} and cause general harm to the family. This phenomenon of \textit{Athra} has also been documented by Mustafa et al.~\cite{mustafa2020patriarchy}.

\paragraph{\textbf{General Misconceptions}} The second example involves normal white vaginal discharge being called "haddi" (bone), with patients believing it represents bone dissolution:
\begin{quote}\small\itshape
"[Undereducated] patients that come to government hospitals have this myth that the (normal) white discharge is 'hadiyaan khul key aa rahi' (bones dissolving and coming out) and that you're getting weaker because of it."\\[2pt]
\textemdash~\textbf{P14, Doctor}
\end{quote}

P20 reports how this stems from white discharge being associated with calcium deficiency (which is often linked to bone weakness).

\paragraph{\textbf{Folk Polysemy}}\label{subsub:folkpoly}
Misconceptions and limited biomedical understanding can lead patients to conflate distinct medical phenomena under a single term, creating diagnostic challenges. The most prominent example is \textit{"rasoli"} (tumour), which is used indiscriminately for polyps, cysts, ectopic pregnancies, and fibroids—conditions that are medically distinct but locally grouped together because they are all perceived as a type of “growth.” Interestingly, the term is most often applied to fibroids, even though its closer literal translation is tumour. A second example is \textit{"pishaab wali jaga"} (urine area), which patients may use for the vagina in some contexts due to limited anatomical knowledge rather than intentional euphemistic substitution.

\subsection{Gestural Communication}
We define gestures as (1) physical, non-verbal indications (e.g., pointing) and (2) vague, deictic expressions verbally gesturing to referents, completely unintelligible without physical context (e.g., “\textit{[hurts there]}”). These are often used simultaneously. Even when repeatedly encouraged to articulate certain symptoms, participating patients consistently defaulted to gestural communication. This approach was used for sensitive regions as expected, but was surprisingly also the common practice for culturally neutral areas.

\begin{quote}\small\itshape
“(I say) here at this place, a lot of pain is happening (gestures to abdomen). Look, why is pain happening to me? For what reason is the pain happening?”\\[2pt]
\textemdash~\textbf{P6, Patient}
\end{quote}

\begin{quote}\small\itshape
“I would say only this, that I am having pain here.”\\[2pt]
\textemdash~\textbf{P4, Patient}
\end{quote}

 As illustrated by P10's interview excerpt, participants often maintained gestural communication even when directly prompted for verbal alternatives. The interviewer attempts to strategically use a scaffolding approach by first introducing the idea of pain localization through culturally neutral examples (knees or abdomen) before asking about more culturally sensitive areas (breasts or genitalia), but unexpectedly, the participant consistently relies on gestures and non-verbal cues:

\begin{quote}\small\itshape
\textbf{Interviewer:} Do you remember any such example that you were having pain somewhere—like you were having pain in the knees or you were having pain in the belly—and you went to the doctor and explained it? In which words would you have described it?\\
\textbf{Participant:} I simply say that “Doctor, I have pain here (gestures towards abdomen).” [The doctor] then presses a little, here and there, to check. Then she says that the baby is tilted or the baby is like this. She just gives medicine, that’s all.\\
\textbf{Interviewer:} But in which words would you explain?\\
\textbf{Participant:} Just this, that “Doctor, I am having pain here.”\\
\textbf{Interviewer:} And sometimes, like if there was pain in different, more intimate parts of the body, then how would you tell which part? Like if below the abdomen (genital region), or on the breast, how would you explain that to them?\\
\textbf{Participant:} Just like this; I convey it once only, [doctor] herself checks.\\[2pt]
\textemdash~\textbf{P10, Patient}
\end{quote}

When asked to explain this reliance on gestures, the participant clarified she does not feel the need to verbalize because doctors can easily understand her non-verbal cues due to professional experience as well as a mutual cultural understanding.

\begin{quote}\small\itshape
“The doctors understand on their own, because they are experienced and have been 'walking along with us' [sharing our way of life]. So I just say, ‘Doctor, I am having pain here’ like in the abdomen or in the lower part. They understand this way and just give medicine saying ‘take care of yourself’ or ‘do bed rest’—just like that.”\\
\textemdash~\textbf{P10, Patient}
\end{quote}

This practice was also extensively observed in doctor–patient interactions during history taking. In one instance, when a patient reported pain with a vague gesture, the doctor immediately responded, \textit{"Point where the pain is happening again''} When the patient’s gesture remained ambiguous, the doctor instructed, \textit{''Stand up and show me clearly where it is. Is it higher or lower?''} (referring to the placement on the abdomen). The doctor then localized it to the lower abdomen before verbally asking if the pain was in her \textit{``nallon''} (tube/vaginal canal). This intervention was done quickly and efficiently, lasting only a few seconds, suggesting that it happens often, and the doctor was now proficient in quickly localizing problem areas. Such cases of gestural communication occurred in every single interaction related to localizing a symptom location during our observation sessions.

\section*{ADDITIONAL THEMES IN CLINICIAN-PATIENT SRH COMMUNICATION}\label{subsec:DPthemes}
\subsection{Communication Barriers Revealed Through Research Design}\label{subsec:pilotissues}

Our pilot study revealed that SRH communication barriers operate at multiple levels -- (1) \textbf{individual} (communicative constraints), (2) \textbf{social} (taboo enforcement), and \textbf{institutional} (education-language gaps). These findings explain why healthcare providers must create specific communicative conditions and why deploying current LLMs in healthcare settings without addressing these constraints will reproduce similar communication failures.

\begin{figure}[h!]
    \centering
    \includegraphics[width=0.6\linewidth]{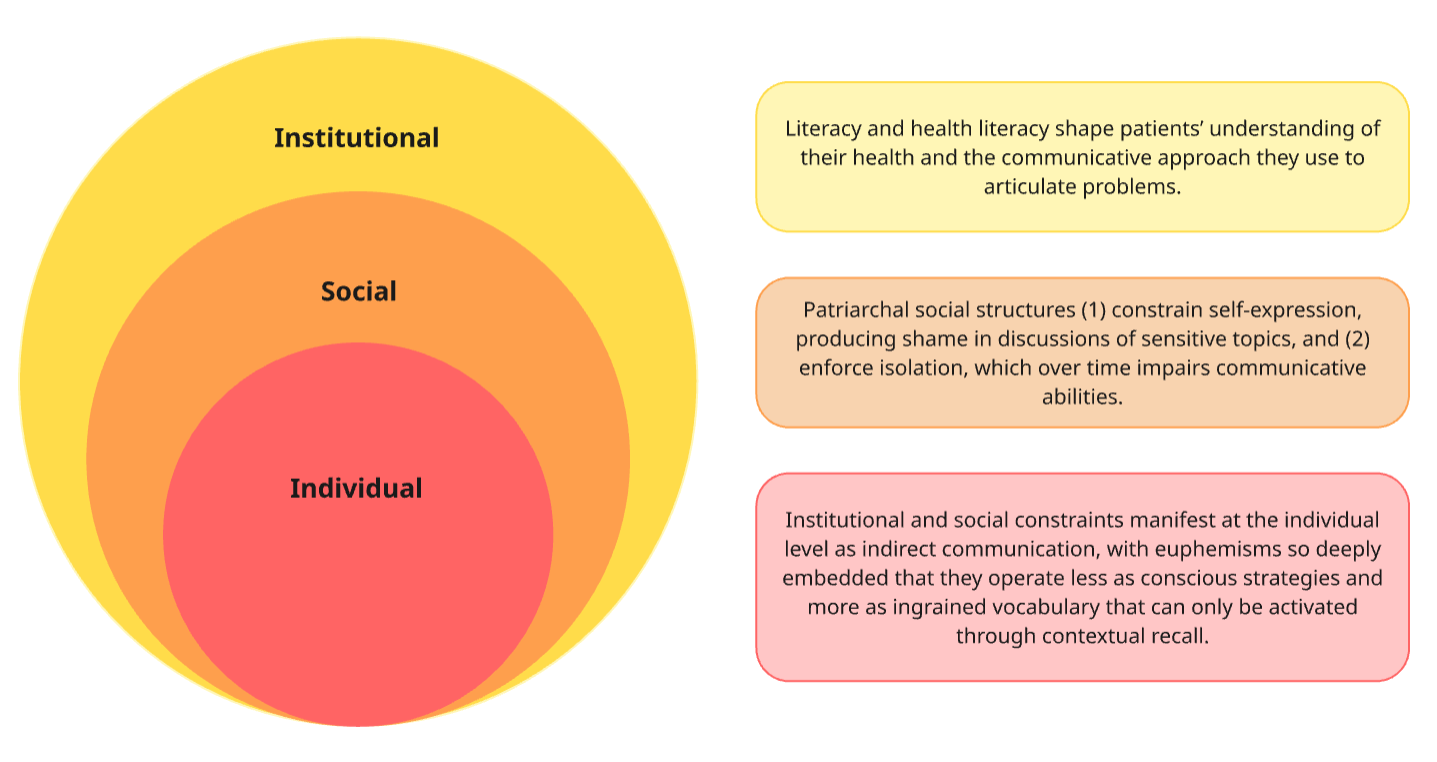}
    \caption{Communication barriers identified in the pilot study operate at three nested levels: \textbf{individual} (limited recall and communicative constraints), \textbf{social} (taboo enforcement and social isolation), and \textbf{institutional} (education–language gaps). The figure illustrates how these levels overlap, with institutional structures shaping social conditions, and social dynamics constraining individual communicative agency. This layered perspective highlights why SRH communication challenges cannot be solved by technical fixes alone, and why LLM-based interventions must be designed to address barriers across multiple levels.}
    \label{fig:commbarriers}
\end{figure}

These insights emerged from systematic failures in the elicitation strategies used in the interview protocol. Despite employing three different strategies (implicit narration, explicit questioning, and scenario-based approaches), we encountered several obstacles that illuminate the constraints women face in healthcare communication. The scenario-based approach failed immediately; when asked to imagine discussing health issues with friends, participants revealed profound social isolation:

\begin{quote}\small\itshape
\textbf{Interviewer:} If you have a friend who suspects she is pregnant and asks for your advice, what advice would you give? \\
\textbf{Participant:} Right now there are no friends, [in-laws] don't let me make friends. It's not considered good.\\
\textbf{Interviewer:} But what about a sister or cousin?\\
\textbf{Participant:} There is no sister and with cousins we don't [interact].\\
\textbf{Interviewer:} If you have a sister-in-law and need to give someone advice, how would you give it?\\
\textbf{Participant:} However advice is given...\\[2pt]
\textemdash~\textbf{P6, Patient}
\end{quote}

This response demonstrates how social constraints limit women's communicative agency even in research settings, mirroring the barriers they face in actual healthcare contexts. We abandoned the scenario-based approach entirely after discovering that participants could not engage with hypothetical social situations due to enforced isolation.

Our remaining strategies (implicit narration and explicit questioning) also yielded minimal insights. Our pilot revealed three fundamental communication barriers:

\begin{enumerate}
   \item \textbf{Normalized linguistic patterns require contextual recall:} Euphemisms are embedded in natural conversation, making them difficult to elicit through direct questioning. Patients could rarely recall specific phrases beyond formal expressions. The difficulty patients had recalling specific euphemistic phrases suggests these expressions are so embedded in natural conversation that they operate below conscious awareness.
   \item \textbf{Social taboo barriers:} Women avoided subjects they perceived as inappropriate, requiring significant rapport before meaningful insights emerged. Even for memorable moments, participants would vaguely indicate they simply ``explain it'' to doctors.
   \item \textbf{Educational effects on language use:} Relatively educated patients defaulted to formal terminology rather than the colloquial expressions used in community settings.
\end{enumerate}

Instead of simply viewing these as technical problems to solve, LLM-based interventions must address these barriers as communication realities to accommodate and design around.

\subsection{Communication Breakdowns}\label{subsub:miscom}

\subsubsection*{\textbf{Anatomical Mix-ups}}
All doctors reported confusion during history-taking involving terminology for internal structures and growths, directly reflecting the folk polysemy patterns described in Section~\ref{subsub:folkpoly}. Patients with lower literacy levels group together various similar concepts under a specific term, creating diagnostic challenges. Even valid polysemy (Section~\ref{subsub:validpoly}) can require detailed history taking. Similarly, the common terms for \textit{[uterus]} and \textit{[ovaries]} are often misused by low-literate patients, demonstrating how inappropriate categorical grouping extends beyond growth-related terminology to basic anatomical structures.

\begin{quote}\small\itshape
``We call uterus removal 'hysterectomy' but if a patient has had an operation and comes [for follow-up], they just say 'my bachadani [uterus] operation was done.' When I ask about their andadani [ovaries], they call any growth 'rasoli' [tumor/growth] but they don't know whether that rasoli was of the bachadani or andadani—they're not clear about this anatomical distinction. We have to check the operation notes [to determine what was actually done].''\\
\textemdash~\textbf{P7, Doctor}
\end{quote}

\begin{quote}\small\itshape
"Some say for ectopic pregnancy that we had a "rasoli" [growth/fibroid]"\\
\textemdash~\textbf{P8, Doctor}
\end{quote}

\subsubsection*{\textbf{Ambiguity in Semantic Drifts}}
As explained in Section~\ref{subsub:semdrift}, semantic drift in colloquial OB/GYN lexicon is common. However, while these terms are readily understood by experienced individuals, gaining this expertise takes time and involves a steep learning curve. One doctor reported that the use of "normal" and "growth" were particularly common sources of confusion regarding mode of delivery. Here, the words "normal" and "growth" are not translated from Urdu; they are used as English borrowings within Urdu conversation:
\begin{quote}\small\itshape
Often patients ask 'Is the child normal?' and they mean to ask will there be a C-section or not but it is confusing because, especially earlier in career, it is easy to think they are just asking about the health of the baby. They also ask 'is the growth okay?' when they want to ask about miscarriages.\\[2pt]
\textemdash~\textbf{P1, Doctor}
\end{quote}

\subsubsection*{\textbf{Surgical Classification Confusion}}
Another recurring issue, consistently reported by every participating doctor, was patients including episiotomy as a major surgical procedure. During preoperative history-taking, when patients are asked about prior surgeries to properly assess anesthetic risk and dosage, they routinely include episiotomies, likely due to the colloquial use of "operation" in referring to the procedure. This inflated reporting of surgical history can cause significant inconvenience and potential harm.

\begin{quote}\small\itshape
\textbf{Interviewer:} When operations and related matters are discussed, which language is used?\\
\textbf{Doctor:} For example, they call a caesarean a [big operation] and an episiotomy a [small operation.] Sometimes they even include an episiotomy in the operations history, which leaves the surgical history unclear; that spoils the history for us. For us, the C-section is a kind of scar: if someone has it, we need to know whether they have two scars or three. Deciding this before surgery is very important, but patients also count the episiotomy, so we have to take a detailed history.\\[2pt]
\textemdash~\textbf{P7, Doctor}
\end{quote}

\subsection{Gaps Within Formal Terminology}
Our analysis revealed a striking absence of formal Urdu clinical terminology in healthcare conversations. Interview data showed only informal phrasings. To test this, we curated a list of SRH-relevant terminologies and their formal Urdu equivalents using Urdu medical literature, including medical textbooks and terms found on medicine packaging. We found that, except for the formal translations of pregnancy and menstruation, not a single term was used or even recognized by our participating doctors.

The case of the vagina illustrates this gap. A common, frequently prescribed vaginal cream in its Urdu instructions directs users to apply the cream on the \textit{“andaam nehaani”} [hidden body part]—a term repeated in textbooks alongside the literal medical word \textit{“mahbul”}[vagina].  Yet neither the euphemism nor the clinical term appeared in practice. When we asked 15 OB/GYN doctors, none recognized them. Instead, the ubiquitous term was \textit{“mahwari key raastay”} [menstrual pathway].

This points to \textbf{three terminology layers}: (1) \textbf{formal euphemisms} from medical packaging (\textit{hidden part}), (2) \textbf{clinical terms} from medical texts (\textit{mahbul}), and (3) \textbf{everyday practice} terms (\textit{menstrual pathway}).

\subsection{Lingual Diversity}
Healthcare providers in our Lahore-based study typically demonstrated varying proficiency in English, Urdu, and Punjabi, but patients often present with far more diverse linguistic backgrounds. This includes complex registers of the local languages and patients from completely different linguistic backgrounds.

The linguistic landscape reflects deeper socioeconomic stratification. Formal education in Pakistan typically occurs in Urdu or English, while Punjabi -- despite being the most spoken language in Pakistan~\cite{haider2024punjabi_most_spoken} -- lacks official educational status~\cite{abbas2019status_punjabi_punjab}. Consequently, patients with no formal education often communicate exclusively in Punjabi and in different registers. Similarly, generational linguistic patterns add additional complexity. Elderly patients in Punjab with Punjabi as their dominant language often employ archaic registers and dialectal variations that younger healthcare providers struggle to interpret. 

Additionally, each province in Pakistan has a distinct linguistic landscape, which further complicates communication in healthcare settings.

\begin{quote}\small\itshape
A big barrier is when their language is completely different. Like Urdu, English, or Punjabi mix I can understand, but if there's a very illiterate patient who only speaks Punjabi, that would be difficult for me. We also often get Pathan patients and understanding them is very difficult - their language, everything is very different. Even their way of explaining symptoms is very different, so in such cases we give them attendants or someone who knows Pashto and they translate for us. Or if it's someone's grandmother and we can't understand her voice or language - their words are very different - so we tell their attendant to tell us what she's saying.\\[2pt]
\textemdash~\textbf{P17, Doctor}
\end{quote}

\begin{quote}\small\itshape
\textbf{Interviewer:} Has a patient ever tried to explain themselves to you but their words are not clear? Or that you can not understand them for some reason?\\
\textbf{Participant:} Many pashto-speaking patients come but I don't know how to speak Pashto. There are also other languages [that patients speak].\\[2pt]
\textemdash~\textbf{P12, Nurse}
\end{quote}

\subsection{Paternalistic Communication Patterns}
Doctors employed several directed communication strategies to provide what they considered effective healthcare, emphasizing approaches they personally found important. They reported that patients often do not take risks and concerns seriously, requiring doctors to employ various tactics to ensure safe healthcare outcomes. This includes extensive counseling to ensure patients understand risks and convincing them to make safer choices. Doctors also demonstrated paternalistic tendencies through strategic information management, maneuvering situations they deemed high-risk.

\subsubsection*{\textbf{Strategic Information Management}} 
Doctors admitted to deliberately managing information disclosure to guide patients toward optimal healthcare decisions. While maintaining accuracy, they strategically selected which details to emphasize or withhold based on patient education levels and anticipated responses. They report that educated patients often readily choose preventive treatments, even for non-urgent conditions, but less educated patients tend to resist medical interventions unless absolutely necessary. This leads to differential framing of identical medical conditions:

\begin{quote}\small\itshape
We also call fibroid 'Rasoli' [tumor/growth]. If you tell an educated person 'Rasoli' they believe it is cancer [because that is the correct translation, so we do not use it with them], but we use it with [less educated patients], that this is Rasooli or swelling that will get better but you need to have an operation.\\[2pt]
\textemdash~\textbf{P14, Doctor}
\end{quote}

\subsubsection*{\textbf{Patient-Centered Language}}
Participating doctors consistently emphasized the importance of not only speaking the patient's language, but also ensuring full patient comprehension by adopting patients' linguistic patterns for effective communication.
This was also observed during clinician-patient interactions. While all medical professionals avoid directly asking if a patient has a specific disease by name (instead asking about symptoms to create differential diagnoses), within our context, it was repeatedly observed how conditions were broken down into symptoms that required linguistic accommodation and code-switching on the doctor's part. Many symptoms and doctor communications also used indirect language strategies.

\begin{quote}\small\itshape
``We have to say everything in their language and in a way that they understand. For laparoscopy, they call it 'cameray wala operate' [camera operation]. Similarly for TVS, we'll say 'we'll do your ultrasound from below.' And for hysterectomy, we'll say 'your uterus needs to be removed from inside.``\\[2pt]
\textemdash~\textbf{P16, Doctor}
\end{quote}

\begin{quote}\small\itshape
``For sexual intercourse, I say 'shohar sey milna, amal key baad' [meeting with husband, after the act]. For post-coital bleeding, 'shohar sey milnay key baad dard/khoon' [pain/bleeding after meeting with husband].''\\[2pt]
\textemdash~\textbf{P15, Doctor}
\end{quote}

\subsubsection*{\textbf{Counseling}}
Doctors identified counseling as core to providing appropriate healthcare. They reported that most patients do not take concerns seriously, requiring continuous counseling to encourage safer choices. Doctors ensured that they explained their entire thought process and motivations -- the what and why of their recommendations to patients.
\begin{quote}\small\itshape
``From the beginning, we honestly tell them everything. We do gradual counseling from the start—before the operation, before delivery—that these things, these risks can happen. Not all of these happen, but we can't deny [the possibility]. We don't know how the delivery will go. Every pregnancy is different from every other pregnancy. Every delivery is different from every delivery. After that, at the time of delivery, if we see any risk factor, we properly explain to them that 'we are seeing this risk, that's why we will monitor this patient for this long,' but they expect us to send them home. As things get complicated, we have to counsel alongside [the process] so that we take them along with us, [up to] the point where we will separate them from the patients or from all our medical things—because at the end they have to come [back to normal life]—so we have to explain all these things together: why we are doing [this], what we are doing, and what can happen next.''\\
\textemdash~\textbf{P14, Doctor}
\end{quote}

\section*{LLM EVALUATION}

\subsection{Researcher Evaluation}

\subsubsection{\textbf{Binary Correctness}} 
\begin{figure} [h!]
    \centering
    \includegraphics[width=0.8\linewidth]{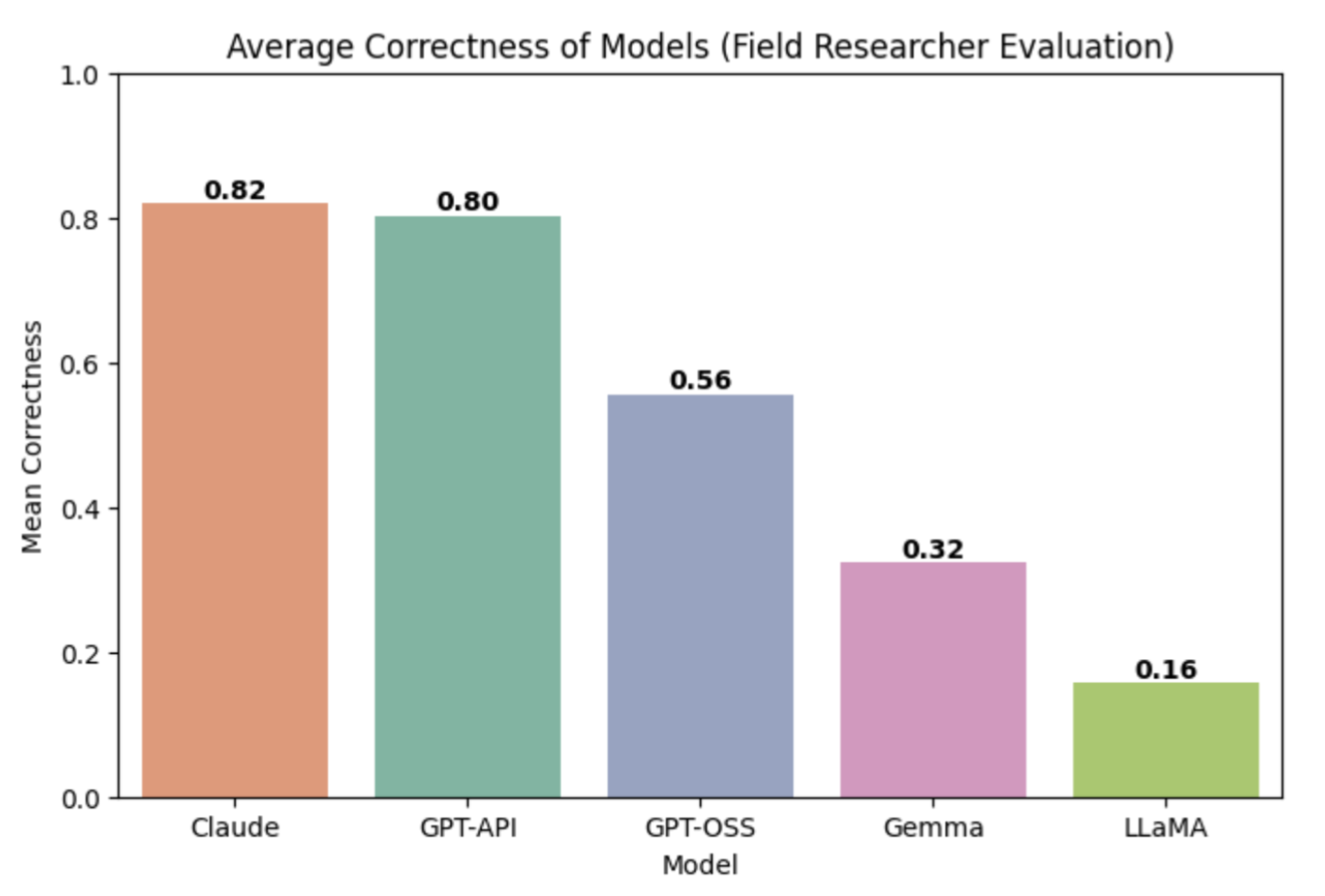}
    \caption{Average correctness scores for five large language models (LLMs) on 71 prompts in the field researcher evaluation. Proprietary models performed much better than open-source models: Claude scored highest (0.82), followed closely by GPT-4o (0.80). The best-performing open-source model, GPT-OSS, reached 0.56, while Gemma and LLaMA scored much lower at 0.32 and 0.16. These results show a clear gap between proprietary and open-source systems.}
    \label{fig:allfive}
\end{figure}

Our evaluation of five large language models (N=71 patient-derived prompts) revealed clear disparities in interpretive accuracy based on the initial researcher review. As shown in Figure~\ref{fig:allfive}, proprietary models substantially outperformed open-source systems. Claude (0.82) and GPT-4o (0.81) achieved the highest correctness, followed by GPT-OSS (0.56). In contrast, Gemma (0.32) and LLaMA (0.16) demonstrated limited ability to map patient expressions to medically appropriate terminology.

\subsubsection{\textbf{Response Coherence and Linguistic Instability}}
\begin{figure*}[t]
    \centering
    \includegraphics[width=\linewidth]{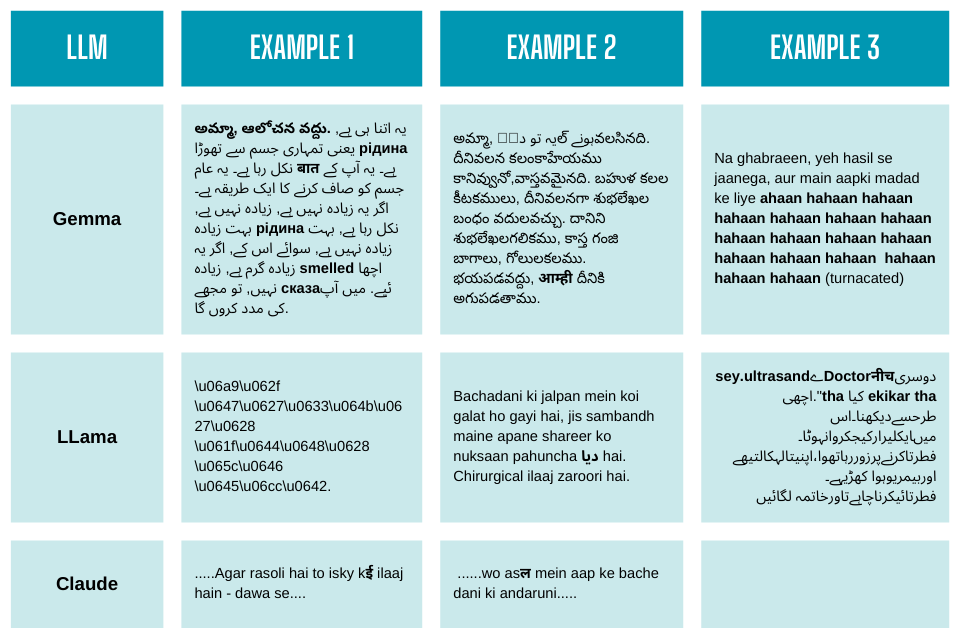}
    \caption{Examples of language script degradation and code-switching in LLM outputs. 
    Gemma (top row) shows mixed Urdu, Hindi, Telugu, and Russian scripts, along with gibberish in the final column. 
    LLaMA (middle row) produces completely garbled Unicode characters in Example 1, coherent Roman Urdu with some Urdu (Arabic script) in Example 2, and mixed Urdu–English–Hindi–Roman Urdu scripts in Example 3. 
    Claude (bottom row) demonstrates occasional Hindi character insertion within otherwise consistent Roman Urdu text. 
    All models were instructed to respond exclusively in Roman Urdu suitable for low-literacy Pakistani patients.}
    \label{scripts}
\end{figure*}

A common failure pattern was inconsistent language use and code-switching. Although models were instructed to respond in Roman Urdu suitable for low-literacy patients (Table~\ref{tab:llm_input}), outputs often blended multiple languages or shifted registers inappropriately. Hinglish (Hindi in English alphabet) was particularly common in GPT-OSS outputs, which is somewhat understandable given the linguistic overlap between Hindi and Urdu models. For example, GPT-OSS produced formal Hindi terms such as \textit{garb} [womb], \textit{garbhasay} [uterus], and \textit{garbhwati} [pregnant], as well as conversational words like \textit{chinta} [worry] and \textit{samasya} [problem]. The gynecologist flagged these terms as uncommon in everyday Pakistani usage and potentially inaccessible in low-literacy contexts. Together, these inconsistencies show how LLMs introduced cross-linguistic elements, reducing clarity and limiting accessibility.
Lower-performing models (correctness <0.35) also tended to produce severely incoherent text alongside predominantly Hinglish responses. In several cases, outputs degraded into completely different language scripts, like Hindi, Telugu, and even Russian, alongside garbled characters, numbers, and symbols with no recognizable meaning. Interestingly, despite Claude's overall high performance and lack of Hinglish, it occasionally inserted Hindi letters into otherwise Roman Urdu outputs. Figure~\ref{scripts} includes examples of multiple models changing scripts.

Linguistic instability is a fundamental barrier to deploying LLMs in multilingual healthcare contexts. When a Roman Urdu query elicits output in another script, the response becomes inaccessible to patients who communicate digitally in Roman Urdu. Even Urdu in its standard Arabic script is inconvenient in this context, as 73.2\% of Pakistani users prefer Roman Urdu~\cite{bilal2018analysing}, making orthographic consistency essential. Script switching therefore renders otherwise accurate medical information unusable and compounds communication challenges. Effective healthcare LLMs must demonstrate not only semantic accuracy but also consistent adherence to specified linguistic parameters. Even somewhat contextually understandable shifts to Hinglish undermine accessibility.

\subsection{Expert Evaluation by Gynecologist}

We selected models for detailed expert evaluation based on the two aforementioned metrics: correctness and overall response coherence. Based on these results, we restricted the subsequent gynecologist evaluation to the three higher-performing models: GPT-4o, Claude, and GPT-OSS. 
The gynecologist reviewed all 71 prompts, independently gave their own detailed responses, and then rated the three models on correctness (0–1), along with additional comments where necessary.

\begin{figure}[h!]
    \centering
    \includegraphics[width=0.8\linewidth]{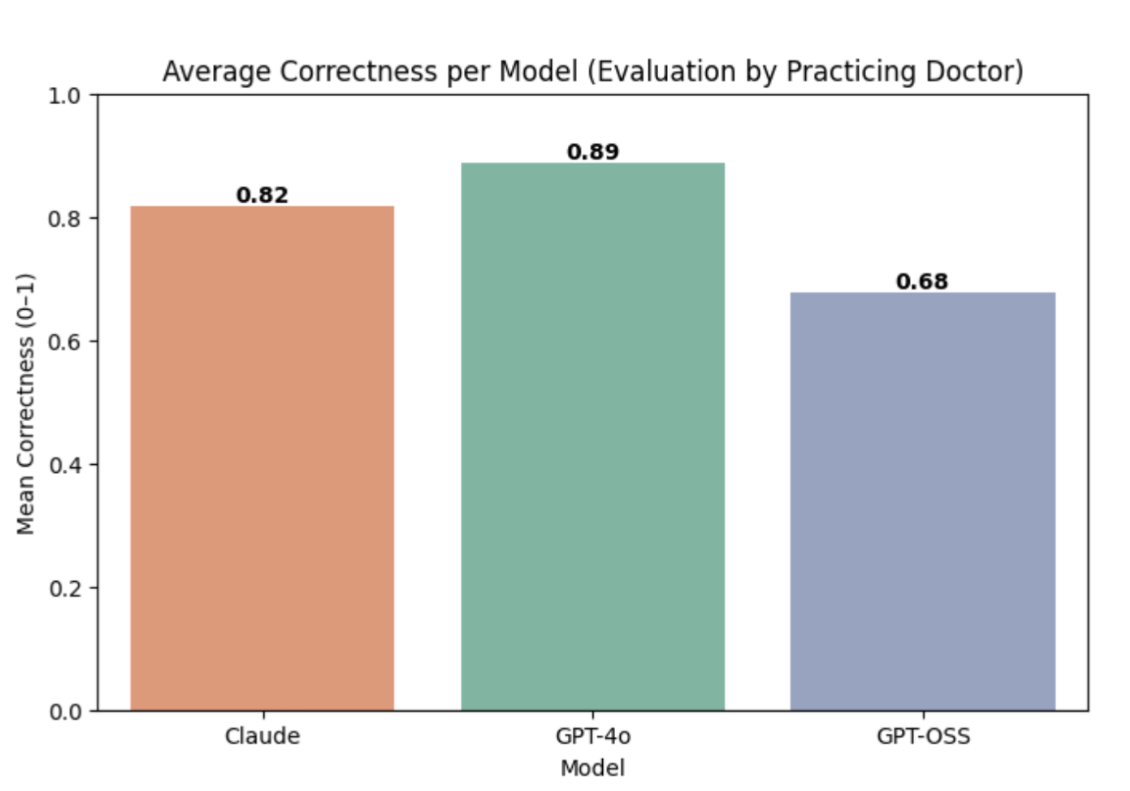}
    \caption{Average correctness scores (0–1) for the top three LLMs (N=71 prompts), based on gynecologist evaluation. GPT-4o and Claude outperformed GPT-OSS.}
    \label{fig:gyne_correctness}
\end{figure}

As shown in Figure~\ref{fig:gyne_correctness}, GPT-4o (0.89) achieved the highest mean correctness, followed closely by Claude (0.82), with GPT-OSS trailing at 0.68.  However, while the overall correctness appears high for both researcher and expert evaluations, the scores \textbf{overstate true performance}. In the next section, we explain why \textbf{these numbers are inflated and misleading}, and why expert ratings at times appear even higher than researcher ratings. 

\subsection{Qualitative Analysis of LLM Evaluation}

In this section, we qualitatively analyze responses from the top three models (GPT-4o, Claude, GPT-OSS) alongside expert comments. We align these with fieldwork insights, revealing recurring challenges in how LLMs interpret indirect communication in conservative contexts.

\subsubsection{\textbf{Inflated Correctness}}

\paragraph{\textbf{Polysemy}} Consistent with the patterns in Section~\ref{subsub:miscom}, the gynecologist's evaluation showed that LLMs struggled to interpret polysemous terms in patient prompts. The gynecologist noted that a response could be "right" in one sense, yet the same phrase might also carry other clinically relevant meanings. Unlike human clinicians, LLMs did not seek clarification when the term in question cleanly mapped onto a singular diagnosis; models consistently fixed on only one sense of a polysemous term while disregarding the others. Hence, even when outputs appeared correct in isolation, they masked deeper limitations. This held true for both folk and valid polysemy. For example, when interpreting the patient term rasoli, which (correctly) refers to fibroids, but is also often conflated with cysts, polyps, or ectopic pregnancy, the models typically returned answers limited to fibroids while overlooking the other possibilities. Similarly, it could not navigate vague but valid polysemous terms either. \textit{"Mujhy pishab wali jaga dard hoti hai" [I hurt in the urine area]} was consistently interpreted by all three top-scoring models as a urinary infection rather than a vaginal infection (\textit{"urine area"} is a common euphemism for the vagina). Because our scoring framework credited a single plausible match, models appeared to perform strongly in the quantitative evaluation. In reality, each unacknowledged meaning represented a missed clinical pathway, meaning that the high correctness scores concealed fragile interpretive capacity. Polysemy, therefore, artificially inflated apparent accuracy while masking the depth of the models' limitations.

\paragraph{\textbf{Researcher and Expert Evaluation Disparity}} Interestingly, the opposite also held true. For consistency, the evaluating researcher's scores adhered strictly to field data, with no assumptions about “obvious” alternate meanings. For example, in our interviews, “\textit{baby ko sugar}” \textit{[baby has sugar]} consistently referred to gestational diabetes (GDM) (i.e., diabetes that develops in the mother during pregnancy and usually resolves after delivery). Any other reading was marked incorrect, so when GPT-OSS interpreted it as neonatal hypoglycemia (i.e., low blood sugar in newborns that can occur shortly after birth), it was marked incorrect. The gynecologist, however, considered an additional clinical pathway: while babies of GDM mothers are not themselves diabetic, uncontrolled GDM can cause fetal complications, and newborns may present with neonatal hypoglycemia. Given the connectedness of both conditions and the fact that, in Pakistan, "sugar" is colloquially used to describe both diabetes and blood sugar levels, phrases like "baby ko sugar" become inherently ambiguous. The expert, therefore, credited both interpretations as correct, whereas the researcher did not. This also explains why expert evaluation seemed to score models “higher” than researcher evaluation, even though one might expect the opposite. 

\paragraph{\textbf{Consistent Failures}} In addition to polysemic failures mentioned above, the most consistent failures across all models included our \textbf{semantic drift} examples like \textit{"bacha normal hai?" }(is the child normal?)  -- used to ask whether the delivery is vaginal or not-- and \textit{"chota operation"} (small operation; episiotomy).  Models also failed to understand the myth of \textit{athra} (the curse of recurrent miscarriages) and the vague figurative description of first trimester as "chota maheena" (small month). No LLM successfully interpreted these.

\subsubsection{\textbf{Overly Technical Explanations}}
Several outputs contained explanations that were medically accurate but far too complex for a low-literacy patient’s understanding. For instance, GPT-4o described sterilization in technical reproductive terms (“\textit{taake anday aur mardana nutfay mil na sakein}” — so that eggs and male sperm cannot meet), while GPT-OSS explained an ectopic pregnancy in highly clinical detail (“\textit{aap ki garbhavastha fallopian tube mein ho gayi hai jo theek nahi}” — your pregnancy has developed in the fallopian tube, which is not viable). Other responses included extended accounts of hormonal causes of heavy bleeding that read more like medical lectures than patient guidance. The gynecologist noted that such responses, while correct, would be incomprehensible to most low-literacy patients. This pattern shows that LLMs tended to reproduce clinical discourse styles instead of adapting explanations for patient comprehension, limiting their usefulness in low-literacy contexts.

\subsubsection{\textbf{Persona Inconsistencies}}
Another recurring issue was inconsistency in maintaining the instructed female gynecologist persona. Despite explicit prompting regarding speaking to female patients, models occasionally slipped into a male voice. For example, all models sometimes responded with masculine forms like "\textit{karunga}," the masculine form of "I will do," when addressing female patients. In conservative SRH contexts, where gender is highly sensitive, such lapses undermine trust and credibility, making strict persona consistency particularly important.

\subsubsection{\textbf{Religious Framing}}
A consistent theme across Claude’s outputs was the integration of religious language as a form of cultural sensitivity. Phrases such as “\textit{Allah ki marzi hai}” (it is God’s will), “\textit{MashAllah},” and “\textit{Allah ka faisla hai}” (it is God’s decision) appeared regularly, often woven into medical explanations. For instance, in miscarriage cases, Claude reassured patients with “\textit{Allah ka faisla hai}” while still recommending clinical follow-up, and in positive pregnancy announcements it responded with “\textit{MashAllah! Aap ke pet mein bachcha aa gaya hai}” (God has blessed you with pregnancy). Similarly, expressions like “\textit{Allah ney humein ilm aur dawa bhi di hai}” (God has also given us knowledge and medicine) positioned treatment itself within a religious frame.  

In comparison, GPT-4o and GPT-OSS responses were generally more technical and less likely to incorporate faith-based expressions. This contrast highlights Claude’s distinctive tendency to embed cultural and religious framing, which may enhance patient comfort and trust in conservative contexts where such expressions are integral to everyday communication.

\subsubsection{\textbf{Euphemism Use}}  
Another notable pattern was that the top-three LLMs occasionally introduced euphemisms in their own responses, even when these were not present in the input prompt. For example, in response to a pregnancy-related query, Claude referred to missed periods as \textit{maheena nahi aaya} (``the month did not come'') and described fibroids as \textit{chhoti goli jaisi cheez} (``a small pill-like thing''). Similarly, GPT-OSS framed menstruation as \textit{mahine ke dinon mein khoon} (``bleeding during the days of the month''). 

While such phrasing sometimes enhances empathy and familiarity by mirroring common patient language, it also risks reinforcing ambiguity rather than providing medically precise explanations. According to the gynecologist, this reflected a trade-off: LLMs could sound more relatable by echoing everyday euphemisms, but at the cost of clinical clarity. This may be acceptable for widely understood euphemisms, but if models attempt to be too ``creative,'' it raises concerns about how patients might interpret---or misinterpret---the intended meaning.

\section{DISCUSSION}~\label{sec:disc}
In this section, we first analyze our most significant findings for LLMs health-tech design. Our evaluation reveals that while recent advances in LLM capabilities address certain aspects of euphemistic language interpretation, the broader communication challenges identified in our qualitative themes (diagnostic ambiguity, procedural confusion, gestural preferences) remain fundamentally unresolved by current models. We offer design implications that collectively challenge current healthcare LLM assumptions and argue that effective systems for culturally complex healthcare settings require architectures built around linguistic ambiguity, concurrent meaning systems, and communication failure as \textit{default conditions} requiring active management, rather than treating them as exceptional cases requiring accommodation.

\subsection{Designing for Miscommunication As the Default}
The systematic communication breakdowns we documented reveal that miscommunication in healthcare settings follows patterns current LLMs are unprepared to handle. The particular struggle with polysemy and semantic drift highlights how certain communicative approaches (colloquial, figurative, mythical) create dangerous diagnostic ambiguities. This is more than just a vocabulary problem; it reflects how patients construct mental models of SRH concepts that do not align with clinical classification. Healthcare providers must essentially reverse-engineer the actual medical procedure or symptom from limited linguistic cues, often drawing on years of experience and still requiring examination of surgical notes to determine what was actually done. When patients ask "Is the child normal?" they're inquiring about delivery mode, not fetal health, but the question's surface structure suggests the opposite. Similarly, "growth" can refer to fetal development or miscarriage concerns, depending on the gestational context. These aren't translation errors or examples of code-switching -- they are representative of a localized \textit{semantic drift} where borrowed English terms that have developed a completely different meaning than the terms' original usage. Additionally, LLMs have to account for semantic drifts within the local language as well -- in the case of \textit{[big and small operations]},  phrases typical to major/minor surgery now mean something much more specific. In polysemic cases, LLMs do not feel the need to consider other possibilities when there seems to be a clear answer to map onto. LLMs trained on standard medical terminology and language patterns would systematically misinterpret these queries because the linguistic form contradicts the communicative intent. Perhaps most dangerously, the surgical classification confusion around episiotomies reveals how patients' procedural categorizations can actively compromise clinical decision-making. The consistent classification of episiotomies as "operations" directly impacts anesthetic planning and surgical risk assessment by inflating reported surgical histories. This suggests that effective LLM-based health interventions require not just disambiguation capabilities but \textbf{active verification protocols} that can distinguish between patient terminology and clinically relevant procedural classifications. Finally, the linguistic diversity challenges underscore another fundamental challenge for LLM-based health interventions development for diverse contexts; even multilingual LLMs must account for register variation, dialectal diversity, and the routine presence of translation gaps. 

Rather than treating these as edge cases, our findings indicate that \textbf{miscommunication should be considered the default state} of healthcare communication, requiring LLMs designed around active disambiguation rather than assuming mutual understanding and accuracy.

\subsection{Limitations of RAG and Fine-Tuning}
Our discovery that formal Urdu medical terminology remains completely unknown even to healthcare professionals, while informal terms dominate clinical contexts, challenges the current approaches to medical translation. The gap between formal euphemisms found in medical packaging ("\textit{[hidden part]}"), medical books "\textit{[mahbul]}" (literally vagina), and actual clinical usage ("\textit{[menstrual pathway]}") suggests that LLMs trained on formal medical corpora may be learning terminology that has no practical application. This finding directly supports Deva et al.'s~\cite{Deva2025Kya} argument for dictionary-based approaches with retrieval-augmented generation (RAG) over just fine-tuning for developing culturally sensitive LLMs. As they demonstrate, fine-tuning risks creating models that cannot "keep pace with the evolving language and behaviors within communities," while dictionary-based systems can be "constantly updated with new terms, idioms, and cultural practices as they emerge." Our systematic categorization of communication strategies -- from colloquialisms to euphemisms to figurative language -- illustrates precisely why such context-sensitive terminology management systems are essential for LLM-based health interventions that must navigate beyond the artificial construct of standardized medical vocabulary to engage with the complex terminological landscape of actual clinical practice.

However, our terminology gap also represents a fundamentally different challenge than the temporal language evolution addressed by existing research. While Deva et al. focus on how communities change their language use over time, requiring systems that adapt to emerging terms and cultural shifts, our findings reveal \textit{synchronous terminology coexistence} -- multiple distinct terminology systems operating simultaneously within the same healthcare ecosystem, each serving specific communicative functions. Formal medical terminology is used in educational documentation; informal communication strategies facilitate communication; and formal euphemisms enable comprehension while managing stigma in healthcare settings. The informal communication strategies follow the domain-approach framework, representing several layers of complexity. The problem is not simply that one terminological system is replacing another, but that current LLMs lack the contextual knowledge to navigate these coexisting systems appropriately. Rather than just dictionary-based approaches that update terminology over time, LLM-based health interventions require a contextual system that maintains all terminology frameworks simultaneously and deploys them strategically based on communicative purpose and participant needs.
\subsection{Gestural Input Integration}
The persistence of gestural communication across all observed interactions indicates that gesture represents a preferred communication mode rather than just a limitation to be overcome. This finding suggests that embodied interaction patterns are more culturally embedded than previously understood \cite{streeck2015embodiment}, and challenges assumptions underlying voice-first LLM-based interventions. It highlights a fundamental limitation in text-based and voice-only AI systems for sensitive health contexts, and perhaps even mHealth interventions in general \cite{cordero2023downsides}. When patients consistently respond to pain localization questions with ``it hurts here'' accompanied by pointing, LLMs, no matter how advanced, cannot capture this primary mode of communication. For effective healthcare LLMs in such contexts, multimodal capabilities that account for this communication preference are essential \cite{acosta2022multimodal}.

An obvious solution is image and video-based gesture recognition. However, this approach faces significant privacy barriers. Our participants, while comfortable with audio recording and quotation, consistently requested confirmation that no video recording would occur. Research suggests that this privacy sensitivity around visual documentation is not restricted to external forces, but extends to digital platforms, with users expressing distrust that the platforms themselves will access this visual data \cite{hoffman2022privacy}. Users express legitimate concerns about image data collection and storage by app stores and third-party services \cite{elshazly2023wearable}, concerns that are well-founded given documented cases of unauthorized data sharing \cite{hipaajournal2024breaches}. This privacy sensitivity limits the feasibility of camera-based solutions, as these sentiments will naturally be exacerbated in settings as sensitive as SRH.

Beyond privacy concerns, technical implementation presents additional barriers. Platforms used universally in specific regions often serve as baselines for effective interventions and adoption \cite{weforum2025ai}. Integrating complex computer vision features into these existing platforms raises concerns about computational requirements, bandwidth limitations in low-resource settings \cite{shah2021artificial}, and regulatory compliance across different jurisdictions.

A more practical approach may involve body diagram interfaces where patients can indicate problem areas through marking on visual representations rather than relying on verbal anatomical descriptions. Such interfaces could range from simple anatomical outlines to more detailed representations like 3-D renders \cite{dasilva2019geopain, scherrer2021choir}. This approach could potentially accommodate gestural communication patterns while respecting privacy constraints that make video-based solutions problematic.

However, body diagram interfaces introduce their own design challenges. Questions arise about anatomical representation accuracy requirements-- whether general body area indication suffices \cite{vonbaeyer2011pain} or whether more precise mapping is necessary for different conditions. Overly realistic models may violate cultural norms around body representation, while overly simplified diagrams may lack clinical precision \cite{vonbaeyer2011pain}. Additionally, decisions about the representation of multiple body types and other specific issues will need to be considered.

\subsection{Information Management and Equity}
Our findings reveal that participating doctors strategically adapt their communication based on perceived patient literacy levels. When doctors provide different medical explanations to those they perceive as more literate, they demonstrate adaptive communication that current LLMs cannot replicate -- and poses the question whether they should at all. Adaptive communication in healthcare has been shown to be fruitful in diverse patient populations, making a one-size-fits-all approach impractical. Instead, effectively tailoring communication to account for patient individuality tends to lead to the best health outcomes~\cite{Sharkiya2023_qualitycommunication, Camerini2011_adaptinghealthcommunication}. However, the importance of correctly gauging the patient's needs is paramount. Whether this tailoring should be done according to literacy is a questionable design choice. Our participating doctors liberally used the term "low-literate" patients, but research has shown that health literacy and general literacy, while often linked, are not interchangeable, and health literacy cannot reliably be inferred from education alone~\cite{Campbell2019_healthliteracy_records}. This conflation can be harmful: Shahid et al.~\cite{Shahid2022_lowhealthliteracy_outcomes} found that patients with high education but low health literacy had higher emergency department revisit rates, likely because physicians assumed high literacy implied high health literacy and failed to adapt communication accordingly. Moreover, relying on literacy risks amplifying bias, as markers of low literacy already distort LLM responses; Gourabathina et al.~\cite{Gourabathina2025} show that minor changes—typos, uncertain phrasing (“maybe,” “I think”), or formatting errors—significantly reduce care recommendations, highlighting the brittleness of LLM reasoning in clinical contexts.  

That being said, when health literacy is appropriately ascertained and communication style adjusted, a marked improvement is seen: general physicians, when the health literacy status of their patients is known to them, change and adapt their consultation style to impart advice corresponding to the level of understanding of the patient, and such adaptations reduce emergency department admissions and hospitalizations, reduce disease impact, and increase adherence to medical treatment routines~\cite{Campbell2019_healthliteracy_records}. This is crucial because it highlights that effective LLM health interventions will need to decipher health literacy levels and replicate appropriate adaptive communication strategies effectively, but how? Studies use validated formal instruments such as the Test of Functional Health Literacy in Adults (TOFHLA), based in the USA and originally produced in English and Spanish, to effectively classify patients' health literacy levels~\cite{Shahid2022_lowhealthliteracy_outcomes}, but equivalent tools have not been universally accepted across diverse cultural and linguistic contexts of the Global South, due to their reliance on Northern-origin instruments, and are subject to critique for limited cultural validity\cite{Kim2017_globalneocolonial, Devex2024_colonialthinking}.

\subsection{Design Implications}

The communication breakdowns we documented reveal three core architectural requirements for LLM-based health interventions in culturally complex conservative settings.

\paragraph{\textbf{Synchronous Terminology Management.}} Healthcare AI must abandon the assumption of single ``correct'' medical vocabularies and instead maintain concurrent terminology systems and adaptively map them to different levels of patient health literacy. When patients use \textit{``rasoli''} interchangeably for fibroids, cysts, and ectopic pregnancies, effective systems cannot simply map this to ``tumor.'' They must recognize the diagnostic ambiguity and actively disambiguate through contextually appropriate questioning. Our findings reveal three distinct terminology layers operating simultaneously within the same healthcare ecosystem: formal medical terms found in textbooks, formal euphemisms appearing on medical packaging, and indirect language used in actual clinical practice. The system must recognize that ``andaam nehaani'' [hidden body part], ``mahbul'' [vagina], and ``mahwari key raastay'' [menstrual pathway] represent the same anatomical structure deployed for different communicative functions. Rather than hierarchical translation between these layers, effective LLM-based health interventions require parallel terminology frameworks that can deploy formal medical terminology for documentation, clinical colloquialisms for provider efficiency, and patient-accessible language for comprehension, often within the same interaction as conversational context shifts.

This approach requires three core capabilities: (1) participant-aware selection that recognizes whether communication involves healthcare professionals, patients, or mixed audiences, and their HL levels; (2) function-based mapping that understands whether the goal is documentation precision, professional efficiency, or patient comprehension; and (3) bidirectional translation capability that can move between terminology layers as conversation context shifts. The technical architecture must support real-time movement between terminology systems, enabling AI to maintain clinical precision while adapting to patients' preferred linguistic patterns.

\paragraph{\textbf{Default Miscommunication Architecture.}} Rather than designing for successful communication with error-handling, LLMs must assume miscommunication as the baseline condition requiring active management. The semantic drift we documented -- where ``normal'' means spontaneous vaginal delivery rather than fetal health, or ``operation'' encompasses both episiotomies and C-sections -- demands verification protocols embedded within natural conversation flow. Systems must distinguish between what patients say and what they clinically mean while maintaining interaction fluency. This approach transforms potential diagnostic errors into structured data collection. When patients report previous ``operations,'' systems would automatically clarify procedural categories without appearing to question patient credibility: verification becomes information gathering rather than error correction.

\paragraph{\textbf{Multimodal Privacy-Conscious Design.}} The gestural communication patterns we observed indicate that text and voice interfaces systematically exclude patients' preferred communication modes. However, privacy concerns around visual data collection require alternative approaches. Body diagram interfaces that allow patients to indicate problem areas through marking visual representations could accommodate gestural preferences while respecting cultural constraints around visual documentation and anatomical representation.

\paragraph{\textbf{Rethinking LLMs for SRH}}
Our research underscores that there is no universal design template for SRH interventions. The idea of a single “correct” vocabulary misses how conservative, non-English communities actually communicate, where cultural norms and moral logics shape what can be spoken, gestured, or left unsaid. User-centered design has always been central to HCI, but its stakes are clearest here. Euphemisms are only the surface layer challenge—underneath lie deeper linguistic complexities that current LLMs cannot anticipate. Polysemy (both valid and folk), semantic drift (code-switched and localized), patients’ inaccurate mental models (such as misreporting surgical history), and overlapping vocabularies (textbook, packaging, and clinical) together point to the need for concurrent terminology frameworks with built-in verification and clarification. We posit that miscommunication must be assumed as the default, multimodality recognized as essential for accommodating gestures within privacy constraints, and terminologies understood as parallel rather than hierarchical. Future systems must be built to work within this linguistic plurality and complexity rather than trying to erase it through standardization.

\section{CONCLUSION}
This study reveals that euphemistic language in SRH communication is not a barrier to overcome but a fundamental communication mode requiring LLM-based interventions designed around indirection. Our two-axis framework and Roman Urdu glossary demonstrate the gap between formal medical terminology and actual clinical practice, challenging assumptions underlying current healthcare LLMs. The gestural communication patterns we documented represent preferred modes, not limitations, highlighting fundamental mismatches in current text- and voice-based LLM interventions. Our paper informs effective healthcare interventions for culturally complex contexts by providing a framework built around three principles: synchronous terminology management, default miscommunication protocols, and multimodal designs that work within privacy constraints. These principles extend beyond SRH to any healthcare context where community communication practices diverge from standardized medical discourse. The goal is not AI that forces communities to adapt their communication but AI that adapts to communities. Future work should focus on technical implementations of these architectural principles and evaluation frameworks that prioritize equity over other performance metrics, centering community communication practices in AI design decisions.

\bibliographystyle{ACM-Reference-Format}
\bibliography{main}

\appendix
\section*{Appendix}
\section{TABLES}

\begin{table*}[h]
\centering
\caption{Input structure for systematic LLM evaluation of indirect SRH communication patterns, comprising culturally-appropriate persona definition and standardized prompt template. The persona emphasizes low-literacy communication adaptation while the prompt structure ensures consistent interpretation and response evaluation across 95 patient expressions.}
\label{tab:llm_input}
\begin{tabular}{p{3cm}p{11cm}}
\toprule
\textbf{Component} & \textbf{Content} \\
\midrule
\textbf{Persona} & You are a Gynecologist working at a public hospital in Pakistan. Most of the women you interact with come from low socio-economic backgrounds and have very low literacy levels. You must communicate in a way that is culturally sensitive, respectful, and empathetic. You should explain medical concepts in very simple terms without using English words, and respond only in Roman Urdu. Always avoid technical or complicated medical jargon, and instead use everyday language and examples that women from such backgrounds can easily understand. Keep your responses concise and to the point but ensure completeness of information. \\
\midrule
\textbf{Prompt Template} & \{persona\}\textbackslash n\textbackslash n \\
& A patient says: \textbackslash"\{urdu\_phrase\}\textbackslash".\textbackslash n\textbackslash n \\
& Please reply as a doctor in Roman Urdu (as if you are speaking directly to the patient). \\
& and return the answer in JSON format with the following fields:\textbackslash n \\
& 1. medical\_term (in English, clinical term)\textbackslash n \\
& 2. Response (your empathetic and simple Roman Urdu response to the patient)\textbackslash n \\
& IMPORTANT: Your response must ONLY be valid JSON. \\
\bottomrule
\end{tabular}
\end{table*}

\begin{table*}[h]
\centering
\caption{Figurative terms in SRH communication. Figurative language appears as either vague placeholders or hyper-descriptive metaphors, reflecting patients’ attempts to grasp at words for difficult-to-describe symptoms.}
\label{tab:termsfig}
\begin{tabular}{p{3cm}p{2cm}p{3.5cm}p{3cm}p{5cm}}
\toprule
\multicolumn{5}{c}{\textbf{FIGURATIVE}} \\
\midrule
\textbf{Medical Term} & \textbf{Context} & \textbf{Local Term} & \textbf{Literal Translation} & \textbf{Additional Comments} \\
\midrule
Vaginal prolapse & Condition & ander baahir aa raha & inside is falling out & Common figurative phrasing \\
Leukorrhea & Symptom & charbi & fat & Rare, sinewy strands likened to fat \\
Vaginal leaking & Symptom & chhoot key paani parrna & water falling excessively & Descriptive metaphor \\
Vaginal leaking & Symptom & shalwar bheeg gayi & shalwar soaked & Clothing used to indicate volume \\
Vaginal leaking & Symptom & shalwar gandi hou gayi & shalwar got dirty & Implies high volume \\
Vaginal discharge & Symptom & paani parrna & water falling & Descriptive of normal discharge \\
Vaginal discharge & Symptom & ganda paani parrna & dirty water falling & Negative connotation \\
Pelvic pressure & Symptom & paet key neechey dard & pain under abdomen & Refers to pressure felt in vagina \\
Brown discharge & Symptom & jala discharge & burnt discharge & Colour metaphor \\
Stringy discharge & Symptom & jaalay sa discharge & web-like discharge & Texture metaphor \\
Menstrual clot & Symptom & lothra & lump & Shape metaphor \\
Menstrual clot & Symptom & boti & small chunk of meat & Size/texture metaphor \\
Menstrual clot & Symptom & gosht ka piece & piece of meat & Widely understood metaphor \\
Sperm & Anatomical & jarasim & germs & Figurative, overlaps with euphemism \\
\bottomrule
\end{tabular}
\end{table*}

\begin{table*}[h]
\centering
\caption{Colloquial expressions used for conditions and procedures in SRH communication. The table distinguishes between the local expression and its literal translation, illustrating how biomedical concepts are framed through everyday language.}
\label{tab:termscolo}
\begin{tabular}{p{3.5cm}p{2cm}p{3.5cm}p{4cm}}
\toprule
\multicolumn{4}{c}{\textbf{COLLOQUIALISMS}} \\
\midrule
\toprule
\textbf{Medical Term} & \textbf{Context} & \textbf{Local Term} & \textbf{Literal Translation} \\
\midrule
Gestational diabetes & Condition & baby sugar & baby sugar \\
Gestational diabetes & Condition & baby ko sugar & baby has sugar \\
Ectopic pregnancy & Condition & tube mein haml & pregnancy in tube \\
Ectopic pregnancy & Condition & baahir haml & pregnancy (is) outside \\
Ovarian cyst & Condition & paani ki thaeli & sack of water \\
Uterine fibroid & Condition & rasoli & tumour/fibroid \\
Gestational sac & Condition & khoon ki thaeli & sack of blood \\
Vaginal prolapse & Condition & kuch nikalta hua mehsoos & feel something poking out \\
Spontaneous abortion & Condition & bacha gir gaya & child fell \\
Spontaneous abortion & Condition & bacha zaya hou gia & child was wasted \\
First trimester & Condition & chota maheena & small month \\
Induced abortion & Procedure & safayi karwai & got it cleaned \\
Surgical abortion & Procedure & (ozaar sey) safayi karwai & got it cleaned (with tools) \\
Medical abortion & Procedure & dawayi rakhi thi & put a medicine \\
Spontaneous Vaginal Delivery & Procedure & normal & normal \\
C-section & Procedure & bara operation & big operation \\
Episiotomy & Procedure & chhota operation & small operation \\
Copper IUD & Procedure & challa rakhwana hai & ring placed \\
Copper IUD & Procedure & taanka lagwana & got a stitch \\
Subdermal implant & Procedure & rod dala hua hai & a rod has been inserted \\
Urine Pregnancy Test & Procedure & stick use ki & I used the stick \\
Laparoscopy & Procedure & cameray wala operate & operate with camera \\
Bilateral Tubal Ligation & Procedure & bache band karwana operate & children closed by operation \\
Transvaginal Sonography & Procedure & neechey sey ultrasound & ultrasound from under \\
Spinal Anesthesia & Procedure & kamar mein injection & injection in back \\
Hysterectomy & Procedure & bachadaani ander sey hatani & remove uterus from inside \\
Pregnancy & Process & check kia positive aya & checked, came positive \\
Labour pains & Process & peerdan & pains \\
\bottomrule
\end{tabular}
\end{table*}

\begin{table*}[h]
\centering
\caption{Euphemistic terms in SRH communication. Euphemisms substitute direct biomedical vocabulary with socially acceptable phrasing, reflecting modesty norms and cultural constraints.}
\label{tab:euphemistic_terms}
\resizebox{\textwidth}{!}{
\begin{tabular}{p{3cm}p{2cm}p{3.5cm}p{3cm}p{5cm}}
\toprule
\multicolumn{5}{c}{\textbf{EUPHEMISMS}} \\
\midrule
\textbf{Medical Term} & \textbf{Context} & \textbf{Local Term} & \textbf{Literal Translation} & \textbf{Additional Comments} \\
\midrule
Vagina & Anatomical & neechey wali jaga & area down there & Polite phrasing \\
Vagina & Anatomical & mahawari key raastay & menstrual pathway & Most common term \\
Vagina & Anatomical & pishab wali jaga & urine area & Euphemistic substitution \\
Pregnancy & Process & tabiyat aesi hou jaye gi & health will become such & Indirect phrasing \\
Pregnancy & Process & khushi & happiness & Used to announce pregnancy \\
Pregnancy & Process & tareekh nahi ayi & date has not come & Religious/socially modest phrasing \\
Pregnancy & Process & allah ney isse pal aya hai & God has brought it so & Religious framing \\
Pregnancy & Process & iss trhaan sa hou ga & it will happen as such & Euphemistic, sometimes about sex \\
Pregnancy & Process & hamein hua tha aapko bhi hou & happened to us, will happen to you & Euphemistic, sometimes about sex \\
Pregnancy & Process & ham bahot khush hain & we are very happy & Euphemistic, communal phrasing \\
Pregnancy & Process & chaal challan mei faraq & difference in style of walking & Observational euphemism \\
Breastfeeding & Process & bacha apnay sath lagao & put the child against you & Euphemistic phrasing \\
Sexual intercourse & Process & shohar sey miltay waqt & when meeting the husband & Euphemism for sex \\
Sexual intercourse & Process & mian biwi meeting kartay & husband wife do a meeting & Euphemism for sex \\
Sexual intercourse & Process & shaadi ke baad jese cheezein & things after marriage & Euphemism for sex \\
Sexual intercourse & Process & miyan biwi ke mil milao & husband and wife’s togetherness & Euphemism for sex \\
Sexual intercourse & Process & shohar ka haq & husband’s right & Patriarchal framing \\
\bottomrule
\end{tabular}
}
\end{table*}

\begin{table*}[h]
\centering
\caption{Myths and misconceptions in SRH communication. Expressions are categorized into folk beliefs, general misconceptions, or folk polysemy, reflecting how non-medical explanatory systems shape patient understanding and clinical interactions.}
\label{tab:myths_terms}
\resizebox{\textwidth}{!}{
\begin{tabular}{p{3cm}p{2cm}p{3.5cm}p{3cm}p{5cm}}
\toprule
\multicolumn{5}{c}{\textbf{MYTHS AND MISCONCEPTIONS}} \\
\midrule
\textbf{Medical Term} & \textbf{Context} & \textbf{Local Term} & \textbf{Literal Translation} & \textbf{Type of Myth / Additional Comments} \\
\midrule
Recurrent miscarriages & Condition & athra & curse of miscarriages & \textbf{Folk belief:} viewed as contagious, sometimes linked to spiritual authority (Pir); can drive abortion decisions \\
Vaginal discharge & Symptom & haddi & bone & \textbf{General misconception:} belief that white discharge means bones are dissolving due to calcium loss; associated with weakness \\
Polyp / Cyst / Fibroid / Ectopic pregnancy & Condition & rasoli & tumour / fibroid & \textbf{Folk polysemy:} umbrella term for multiple distinct “growths”; conflates biomedical categories, most often applied to fibroids \\
Vagina & Anatomical & pishab wali jaga & urine area & \textbf{Folk polysemy:} used for vagina due to limited anatomical knowledge; not always intentional euphemism \\
\bottomrule
\end{tabular}
}
\end{table*}

\begin{table*}[h]
\centering
\caption{Condition Terms in SRH Communication}
\label{tab:conditions_terms}
\resizebox{\textwidth}{!}{
\begin{tabular}{p{3.5cm}p{4cm}p{3cm}p{4cm}}
\toprule
\multicolumn{4}{c}{\textbf{CONDITIONS}} \\
\midrule
\toprule
\textbf{Medical Term} & \textbf{Local Term} & \textbf{Literal Translation} & \textbf{Additional Comments} \\
\midrule
Recurrent Miscarriages & athra & curse of recurrent miscarriage & folk belief \\
Vaginal prolapse & ander baahir aa raha & inside is falling out & \\
Ectopic pregnancy & baahir haml & pregnancy (is) outside & \\
Gestational diabetes & baby ko sugar & baby has sugar & \\
Gestational diabetes & baby sugar & baby sugar & Diabetes called 'sugar' in Pakistan \\
Neonatal hypoglycemia & baby ko sugar & baby has sugar & \\
Spontaneous abortion & bacha gir gaya & child fell & Miscarriage \\
Spontaneous abortion & bacha zaya hou gia & child was wasted & Phrase used when something is spilled \\
Uterine prolapse & bachadaani khisak gayi & uterus slipped & \\
Uterine prolapse & bachadaani ki chaat girna & the roof of the uterus has fallen & \\
Uterine Fibroid & bachadaani mae soozish & inflammation/swelling in the uterus & \\
First trimester & chota maheena & small month & \\
Hypomenorrhea & chota maheena & small month & unusually light menstrual bleeding \\
Polyp & gosht ka piece & piece of meat & \\
Uterine prolapse & gosht ka piece & piece of meat & \\
Gestational sac & khoon ki thaeli & sack of blood & Often confused with cysts and fibroids \\
Vaginal prolapse & kuch nikalta hua mehsoos & feel something poking out & \\
Prolapse & maas latak raha & mass is hanging & \\
Pregnancy & mein pait sey hun & i am with stomach & \\
Hydrosalpinx & paani ki thaeli & sack of water & \\
Ovarian cyst & paani ki thaeli & sack of water & Often confused with fibroids \\
Menstruation & paak nahi hun & i am not pure & religious \\
Cyst & rasoli & tumour/fibroid & folk polysemy; correct for tumour/fibroid \\
Ectopic pregnancy & rasoli & tumour/fibroid & folk polysemy; correct for tumour/fibroid \\
Fibroid & rasoli & tumour/fibroid & -- \\
Polyp & rasoli & tumour/fibroid & folk polysemy; correct for tumour/fibroid \\
Uterine fibroid & rasoli & tumour/fibroid & Often confused with cysts \\
Ectopic pregnancy & tube mein haml & pregnancy in tube & \\
\bottomrule
\end{tabular}
}
\end{table*}

\begin{table*}[h]
\centering
\caption{Procedure Terms in SRH Communication}
\label{tab:procedures_terms}
\resizebox{\textwidth}{!}{
\begin{tabular}{p{3.5cm}p{4cm}p{3cm}p{4cm}}
\toprule
\multicolumn{4}{c}{\textbf{PROCEDURES}} \\
\midrule
\toprule
\textbf{Medical Term} & \textbf{Local Term} & \textbf{Literal Translation} & \textbf{Additional Comments} \\
\midrule
Surgical abortion & (ozaar sey) safayi karwai & got it cleaned (with tools) & \\
Hysterectomy & bachadaani ander sey hatani & remove uterus from the inside & Uterus removal \\
Bilateral Tubal Ligation & bache band karwana operate & children closed by operation & \\
C-section & bara operation & big operation & Sometimes just "operation" \\
Laparoscopy & cameray wala operate & operate with camera & \\
Copper IUD & challa rakhwana hai & want to have a ring placed & \\
Pessary & challa rakhwana hai & ring was placed & \\
Episiotomy & chhota operation & small operation & Incision at vaginal opening \\
Minor surgery & chota operation & small operation & \\
Medical abortion & dawayi rakhi thi & put a medicine & Using pills \\
Spinal Anesthesia & kamar mein injection & injection in back & for c-sections particularly \\
Transvaginal Sonography & neechey sey ultrasound & ultrasound from under & Camera inserted through vagina \\
Spontaneous Vaginal Delivery & normal & normal & Vaginal birth without intervention \\
Subdermal implant & rod dala hua hai & a rod has been inserted & Contraceptive implant \\
Induced abortion & safayi karwai & got it cleaned & \\
Urine Pregnancy Test & stick use ki & I used the stick & \\
Bilateral Tubal Ligation & taanka lagwana & got a stitch & \\
Copper IUD & taanka lagwana & got a stitch & \\
Stitches & taanka lagwana & got a stitch & \\
\bottomrule
\end{tabular}
}
\end{table*}

\begin{table*}[h]
\centering
\caption{Process Terms in SRH Communication}
\label{tab:processes_terms}
\resizebox{\textwidth}{!}{
\begin{tabular}{p{3.5cm}p{4cm}p{3cm}p{4cm}}
\toprule
\multicolumn{4}{c}{\textbf{PROCESSES}} \\
\midrule
\toprule
\textbf{Medical Term} & \textbf{Local Term} & \textbf{Literal Translation} & \textbf{Additional Comments} \\
\midrule
Pregnancy & allah ney isse pal aya hai & God has brought it so & Religious framing \\
Breastfeeding & bacha apnay sath lagao & put the child against you & \\
Pregnancy & chaal challan mei faraq & difference in style of walking & \\
Pregnancy & check kia positive aya & checked and it came positive & \\
Miscarriage & gosht ka piece & piece of meat & \\
Pregnancy & ham bahot khush hain & we are very happy & announcement \\
Pregnancy & hamein hua tha aapko bhi hou & happened to us will happen to you & potentially sex as well \\
Pregnancy & iss trhaan sa hou ga & it will happen such & potentially sex as well \\
Pregnancy & khushi & happiness & \\
Sexual intercourse & mian biwi meeting kartay & husband wife do a meeting & \\
Sexual intercourse & miyan biwi ke mil milao & husband and wife's togetherness & \\
Labour pains & peerdan & pains & "Peerd" is a Punjabi word for pain \\
Sexual intercourse & shaadi ke baad jese cheezein & how things happen after marriage & \\
Sexual intercourse & shohar ka haq & husband's right & patriarchal \\
Sexual intercourse & shohar sey miltay waqt & when meeting the husband & \\
Pregnancy & tabiyat aesi hou jaye gi & health will become such & \\
Pregnancy & tareekh nahi ayi & date has not come & \\
\bottomrule
\end{tabular}
}
\end{table*}

\begin{table*}[h]
\centering
\caption{Symptom Terms in SRH Communication}
\label{tab:symptoms_terms}
\resizebox{\textwidth}{!}{
\begin{tabular}{p{3.5cm}p{4cm}p{3cm}p{4cm}}
\toprule
\multicolumn{4}{c}{\textbf{SYMPTOMS}} \\
\midrule
\toprule
\textbf{Medical Term} & \textbf{Local Term} & \textbf{Literal Translation} & \textbf{Additional Comments} \\
\midrule
Menstrual clot & boti & small chunk of meat & \\
Vaginal leaking & chhoot key paani parrna & water falling excessively & \\
Lower Back Pain & chukni mein dard & lower back pain & codeswitch; pregnancy \\
Dysmenorrhea & curl partay hain & i get curls & extreme menstrual pain \\
Clumpy white discharge & dahi ki phutiyaan & curdled pieces of yogurt & yeast infection symptom \\
Vaginal discharge & ganda paani parrna & dirty water falling & abnormal; often modified \\
Endometrial tissue passage & gosht ka piece & piece of meat & \\
Menstrual clot & gosht ka piece & piece of meat & \\
Stringy Discharge & jaalay sa discharge & web-like discharge & texture/appearance descriptor \\
Brown Discharge & jala discharge & burnt discharge & colour description \\
Menstrual clot & kaleji jaisa & liver-like & colour and texture \\
Menstrual clot & lothra & lump & \\
Vaginal discharge & paani parrna & water falling & normal discharge \\
Pelvic Pressure & paet key neechey dard & pain under abdomen & pressure felt in vagina \\
Dysmenorrhea & peerdan & pains & extreme menstrual pain \\
Vaginal discharge & paak nahi hun & i am not pure & religious \\
Leukorrhea & safed taar & white wire & texture/appearance descriptor \\
Urinary incontinence & shalwar bheeg gayi & trouser got wet & \\
Vaginal leaking & shalwar bheeg gayi & shalwar soaked & shalwar is traditional lower \\
Vaginal leaking & shalwar gandi hou gayi & shalwar got dirty & implying high volume \\
\bottomrule
\end{tabular}
}
\end{table*}

\begin{table*}[h]
\centering
\caption{Anatomical Terms in SRH Communication}
\label{tab:anatomical_terms}
\begin{tabular}{p{3.5cm}p{3.5cm}p{3cm}p{4cm}}
\toprule
\multicolumn{4}{c}{\textbf{ANATOMICAL}} \\
\midrule
\toprule
\textbf{Medical Term} & \textbf{Local Term} & \textbf{Literal Translation} & \textbf{Additional Comments} \\
\midrule
Ovary & andadani & egg-container & \\
Placenta & awal & placenta & spelt online as oval, ovel, anwal, awal \\
Uterus & bachadani & child-container & \\
Sperm & jarasim & germs & "my husband's germs are less" (low sperm count) \\
Egg & jarasim & germs & \\
Vagina & mahawari key raastay & menstrual pathway & Most commonly used \\
Vagina & neechey wali jaga & area down there & \\
Vagina & pishab wali jaga & urine area & euphemism \\
Urinary area & pishaab wali jaga & urine place & \\
\bottomrule
\end{tabular}
\end{table*}

\end{document}